\RequirePackage{xcolor}
\documentclass[journal,twoside,web]{IEEEtran}          
\usepackage{etoolbox}
\makeatletter
\@ifundefined{color@begingroup}%
{\let\color@begingroup\relax
	\let\color@endgroup\relax}{}%
\def\fix@ieeecolor@hbox#1{%
	\hbox{\color@begingroup#1\color@endgroup}}
\patchcmd\@makecaption{\hbox}{\fix@ieeecolor@hbox}{}{\FAILED}
\patchcmd\@makecaption{\hbox}{\fix@ieeecolor@hbox}{}{\FAILED}

\def\journalname{IEEE TRANSACTIONS ON COMPUTATIONAL IMAGING}

\usepackage{cite}
\usepackage{amsmath,amssymb,amsfonts}
\usepackage{algorithmic}
\usepackage[pdftex]{graphicx}
\graphicspath{{figures/}}
\usepackage{textcomp}
\usepackage{comment}
\usepackage[english]{babel}
\usepackage[utf8]{inputenc}
\usepackage{mathrsfs}
\usepackage[plain,english]{fancyref}
\usepackage{bm, bbold}
\usepackage{physics} 
\usepackage{scalerel}
\renewcommand{\vec}[1]{{\bm{#1}}}
\renewcommand{\op}{\mathbf}
\newcommand\unity{\mathbf{I}}
\newcommand\C{\mathbb{C}}
\newcommand\R{\mathbb{R}}

\DeclareMathOperator*{\argmin}{arg\,min}
\newcommand{\trans}{{\mkern-1.5mu\scaleobj{0.7}{\mathsf{T}}}}
\newcommand{\herm}{{\scaleobj{0.7}{\mathsf{H}}}}

\usepackage{tikz}

\usepackage[small,bf]{caption}

\usepackage{sidecap}
\sidecaptionvpos{figure}{t}
\usepackage{subcaption} 
\usepackage{float} 
\usepackage{hyperref}
\usepackage{textcomp}
\usepackage{xfrac}
\usepackage{array}
\newcolumntype{C}[1]{>{\centering}m{#1}}
\usepackage[strings]{underscore}

\makeatletter
\newcommand\footnoteref[1]{\protected@xdef\@thefnmark{\ref{#1}}\@footnotemark}
\makeatother

\usetikzlibrary{svg.path}
\usepackage{scalerel}

\definecolor{orcidlogocol}{HTML}{A6CE39}
\tikzset{
	orcidlogo/.pic={
		\fill[orcidlogocol] svg{M256,128c0,70.7-57.3,128-128,128C57.3,256,0,198.7,0,128C0,57.3,57.3,0,128,0C198.7,0,256,57.3,256,128z};
		\fill[white] svg{M86.3,186.2H70.9V79.1h15.4v48.4V186.2z}
		svg{M108.9,79.1h41.6c39.6,0,57,28.3,57,53.6c0,27.5-21.5,53.6-56.8,53.6h-41.8V79.1z M124.3,172.4h24.5c34.9,0,42.9-26.5,42.9-39.7c0-21.5-13.7-39.7-43.7-39.7h-23.7V172.4z}
		svg{M88.7,56.8c0,5.5-4.5,10.1-10.1,10.1c-5.6,0-10.1-4.6-10.1-10.1c0-5.6,4.5-10.1,10.1-10.1C84.2,46.7,88.7,51.3,88.7,56.8z};
	}
}

\newcommand\orcidicon[1]{\href{https://orcid.org/#1}{\mbox{\scalerel*{
				\begin{tikzpicture}[yscale=-1,transform shape]
					\pic{orcidlogo};
				\end{tikzpicture}
			}{|}}}}

\newcommand{\onecol}{\linewidth}

\def\BibTeX{{\rm B\kern-.05em{\sc i\kern-.025em b}\kern-.08em
		T\kern-.1667em\lower.7ex\hbox{E}\kern-.125emX}}

\begin{document}
	\begin{center}
		\begin{minipage}{18cm}
			\centering
			
			\Large{PINQI: An End-to-End Physics-Informed Approach to Learned Quantitative MRI Reconstruction}\\ \vspace{0.2cm}
			\large{Felix F. Zimmermann \orcidicon{0000-0002-0862-8973}, Christoph Kolbitsch \orcidicon{0000-0002-4355-8368},  Patrick Schuenke \orcidicon{0000-0002-3179-4830}, and Andreas Kofler \orcidicon{0000-0001-9169-2572}}\\ \vspace{0.8cm}
			
			\raggedright{
				This paper has been accepted for publication in IEEE Transactions on Computational Imaging.
				This is the author's version of an article that has, or will be, published in this journal or conference.
				Changes were, or will be, made to this version by the publisher before publication.\\ \vspace{0.2cm}
				
				\textbf{DOI:} \href{https://doi.org/10.1109/TCI.2024.3388869}{10.1109/TCI.2024.3388869} \\
				\textbf{IEEE Xplore}: \url{https://ieeexplore.ieee.org/abstract/document/10499888/}\\ \vspace{0.5cm}
				
				\textbf{Please cite this paper as:}\\
				\textit{Zimmermann, FF., Kolbitsch, C., Schuenke, P. and Kofler, A., 2024. PINQI: an end-to-end physics-informed approach to learned quantitative MRI reconstruction. IEEE Transactions on Computational Imaging. 10.1109/TCI.2024.3388869}\\ \vspace{0.5cm}
				
				\textbf{Corresponding Author:} Felix Frederik Zimmermann \\
				\textit{felix.zimmermann@ptb.de, zimmf@physik.tu-berlin.de} \\
				\vspace{0.5cm}
				
				\textbf{Acknowledgments:} This work was supported in part by the Metrology for Artificial Intelligence for Medicine (M4AIM) project that is funded by the German Federal Ministry for Economic Affairs and Climate Action (BMWi) in the framework of the QI-Digital initiative. This work was funded in part by the Deutsche Forschungsgemeinschaft (DFG, German Research Foundation), Grant 372486779.
			}
		\end{minipage}
	\end{center}
	\thispagestyle{empty}
	\clearpage
	\setcounter{page}{1}
	
	\title{PINQI: An End-to-End Physics-Informed Approach to Learned Quantitative MRI Reconstruction }
	\author{{Felix F. Zimmermann \orcidicon{0000-0002-0862-8973}, Christoph Kolbitsch \orcidicon{0000-0002-4355-8368},  Patrick Schuenke \orcidicon{0000-0002-3179-4830}, and Andreas Kofler \orcidicon{0000-0001-9169-2572}}
		\thanks{This paragraph of the first footnote will contain the date on which you submitted your paper for review}
		\thanks{This work was supported in part by the Metrology for Artificial Intelligence for Medicine (M4AIM) project that is funded by the German Federal Ministry for Economic Affairs and Climate Action (BMWi) in the framework of the QI-Digital initiative. This work was funded in part by the Deutsche Forschungsgemeinschaft (DFG, German Research Foundation), Grant 372486779.}
		\thanks{F. F. Zimmermann (e-mail:felix.zimmermann@ptb.de), C. Kolbitsch, P. Schuenke, and A. Kofler are with Physikalisch-Technische Bundesanstalt (PTB), Braunschweig and Berlin, Germany}}
	
	\maketitle
	
	\begin{abstract}
		Quantitative Magnetic Resonance Imaging (qMRI) enables the reproducible measurement of biophysical parameters in tissue. The challenge lies in solving a nonlinear, ill-posed inverse problem to obtain the desired tissue parameter maps from acquired raw data. While various learned and non-learned approaches have been proposed, the existing learned methods fail to fully exploit the prior knowledge about the underlying MR physics, i.e.\ the signal model and the acquisition model. In this paper, we propose PINQI, a novel qMRI reconstruction method that integrates the knowledge about the signal, acquisition model, and learned regularization into a single end-to-end trainable neural network. Our approach is based on unrolled alternating optimization, utilizing differentiable optimization blocks to solve inner linear and non-linear optimization tasks, as well as convolutional layers for regularization of the intermediate qualitative images and parameter maps. This design enables PINQI to leverage the advantages of both the signal model and learned regularization. We evaluate the performance of our proposed network by comparing it with recently published approaches in the context of highly undersampled $T_1$-mapping, using both a simulated brain dataset, as well as real scanner data acquired from a physical phantom and in-vivo data from healthy volunteers. The results demonstrate the superiority of our proposed solution over existing methods and highlight the effectiveness of our method in real-world scenarios.
	\end{abstract}
	\begin{IEEEkeywords}
		differentiable optimization, learned regularization, neural network, $T_1$-mapping, unrolled optimization, quantitative magnetic resonance imaging.
	\end{IEEEkeywords}

\newcommand{\rev}[1]{#1} 

\section{Introduction}
Magnetic Resonance Imaging (MRI) is a well-known method and an indispensable tool in clinical practice. However, the most commonly used protocols are qualitative, where the contrast in the images is determined by a mixture of different tissue parameters and acquisition properties. 
To overcome this issue, quantitative MRI (qMRI) techniques such as MR relaxometry have been proposed, which allow for the quantification of specific (bio)physical parameters of tissue such as spin-lattice ($T_1$) and spin-spin relaxation times ($T_2$). These quantitative measurements allow greater comparability between different devices at different sites and can be used to create more specific clinical guidelines. Typically, in qMRI series of qualitative images with different acquisition parameters are recorded and a non-linear inverse problem is solved to obtain the tissue parameters. The clinical application of MRI and especially qMRI is challenging due to the relatively long measurement times. Hence, the acquired data is typically undersampled in Fourier space (k-space) to reduce the scan time at the cost of making the problem more ill-posed. This leads to artifacts that need to be accounted for by adopting appropriate regularization methods. Both for the reconstruction of purely qualitative images, as well as quantitive parameter maps, different approaches utilizing parallel data recording with multiple receiver coils \cite{sense},  compressed sensing, and varying regularization schemes \cite{lustigcs, doneva2010,wang2018,zhao2014}, model-based reconstruction \cite{Block2007,tran2013,becker2019}, and combinations thereof have been proposed. More recently, neural networks were introduced as learned image reconstructions and as data-driven regularization methods \rev{\cite{sriram,modl,dopamine,deept1,correct}}, or for mapping reconstructed images to quantitative maps  \cite{mantis,myomapnet,drone}. 
\begin{figure}
    \centering
    \includegraphics[width=\onecol]{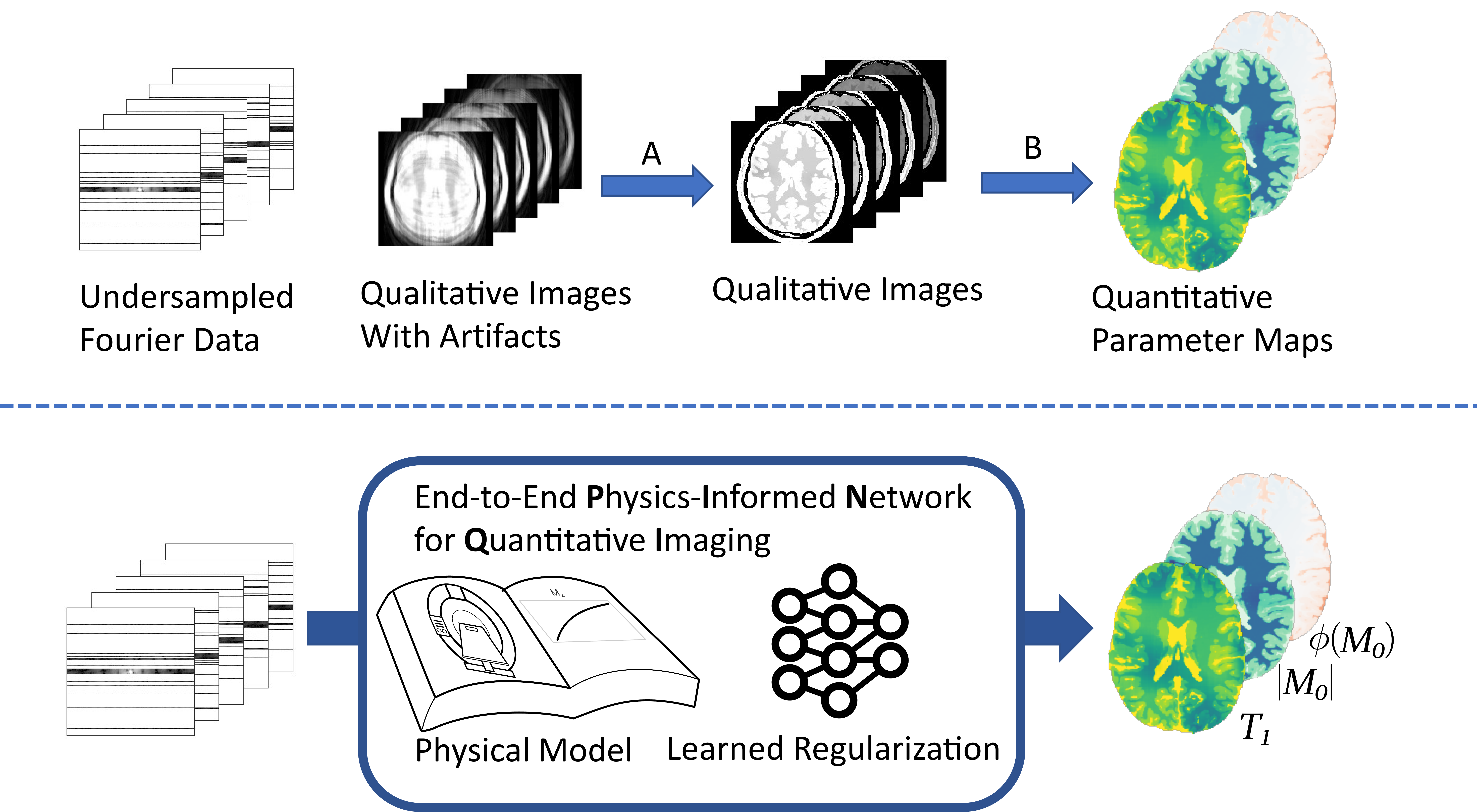}
    \caption{The problem to be solved in quantitative MRI is obtaining maps of physical parameters from undersampled measurements in Fourier space. Most previous methods consider the two steps of A) reconstructing artifact-free qualitative images and B) obtaining the parameter maps as two disjoint and independent steps (see text for details). We propose a novel end-to-end method making use of prior knowledge about the physics of data acquisition and learned regularization.}
    \label{fig:idea}
\end{figure}

The existing data-driven approaches to solve the qMRI inverse problem, i.e.\ to obtain the parameter maps from the acquired k-space data, can be broadly categorized as follows. The first type of approach splits the problem into two disjoint tasks: A) image reconstruction and B) parameter regression (see \Fref{fig:idea}). The second type jointly solves both tasks to obtain the parameter maps directly from the k-space data. For the first task of qualitative image reconstruction, data-driven methods have made great progress: State-of-the-art results can be achieved by incorporating deep learning into an unrolled model-based reconstruction by iteratively applying a neural network as a form of learned regularization as well as enforcing consistency with the recorded k-space data and the linear signal model \cite{sriram,modl,schlemper,crnn,admmnet}. The second task, parameter regression, is then either carried out by classical regression using non-linear solvers \cite{sasha,bojorquez2017,bfgs,numopt}, dictionary matching \cite{mrf}, or by learned methods. Within the latter, a further distinction can be made between pixelwise mappings, such as MyoMapNet for $T_1$-mapping \cite{myomapnet,drone} and methods using the information of multiple pixels via convolutional neural networks (CNNs), thereby implicitly learning a suitable spatial regularization \cite{mantis,deept1,supermap}. 
 MANTIS \cite{mantis}, for example, uses a UNet \cite{ronneberger2015} to predict the parameter maps (originally $T_2$ maps) from qualitative magnitude images with severe undersampling artifacts. DeepT1\cite{deept1}, proposed for myocardial $T_1$-mapping, consists of two separate data-driven parts. First, an iterative reconstruction using recurrent CNNs \cite{crnn} and data-consistency layers for image reconstruction is trained to reconstruct artifact-free qualitative images. Second, a UNet is trained to predict the quantitative maps from the magnitude of the reconstructed images. \rev{A similar approach is used in the recently proposed CoRRECT \cite{correct} for motion-corrected $R_2^*$-mapping. It also uses an unrolled, CNN-regularized image reconstruction, followed by a parameter mapping UNet, with the major difference that both parts are trained simultaneously end-to-end.} However, most learning-based methods entirely ignore prior knowledge about the physics of the signal and acquisition model relating the parameters with the quantitative images or use it only during training as part of a loss function \cite{mantis,supermap,hammernikreview}. Few learned methods are trained in an end-to-end manner, i.e.\ from k-space data to parameter maps, and fully incorporate the physical model at inference time. 

  PGD-Net \cite{deepunrolling} unrolls a proximal gradient descent scheme with an approximated signal function, which is implemented as a pre-trained neural network and used as a differentiable proxy of the true MR fingerprinting signal function. 
  DOPAMINE \cite{dopamine} unrolls a first-order gradient descent scheme that includes an implicitly learned regularizer where small residual CNNs operating on the different parameter maps are used to learn the gradient of a regularizer. Neither employs image-space regularization nor makes use of the particular form of the forward model and both only use shallow networks.
  
 Combining the knowledge about the physics of the acquisition model by data-consistency layers with deep learning regularization has greatly improved qualitative image reconstruction \cite{modl,schlemper,kofler}. Thus, we want to investigate a novel approach to incorporate the full knowledge of the physics of the signal model into a learned qMRI reconstruction, while employing learned regularization in image- and parameter-space.

\subsection{Our Contributions}
Our main contributions to the field of deep-learning-based quantitative imaging consist of
\begin{itemize}
    \item Introduction of non-linear optimization as a differentiable layer into qMRI reconstruction.
    \item A novel and general end-to-end \textbf{P}hysics-\textbf{I}nformed \textbf{N}etwork for \textbf{Q}uantitative \textbf{I}maging, \textit{PINQI}, based on unrolled half quadratic splitting and differentiable optimization.
    \item Validation of the proposed approach on the task of $T_1$-mapping. We show the transferability of the network trained solely on synthetic data to in-vivo measurements.
\end{itemize}
We also demonstrate the usage of implicit differentiation for efficient evaluation of gradients with respect to all inputs of the commonly data-consistency layers of linear inverse problems as well as the non-linear optimization layer.

\section{Methology}
First, we introduce the notation and formulate the inverse problem to be solved in quantitative imaging. Next, we  give a short introduction to differentiable optimization. We present implicit differentiation as a technique to obtain the Jacobian of solutions of optimization problems with respect to the parameters of the problems. Finally, we introduce our proposed PINQI solution to the quantitative imaging problem using differentiable optimization.
\subsection{Problem Statement}
By $\vec{p}(\vec{r})=\left[p_1(\vec{r}), \ldots, p_{N_P}(\vec{r})\right]^\trans$ we denote for each location $\vec{r}$ the $N_P$ relevant physical parameters to be determined, such as relaxation times. For notational brevity, we will write $\vec{p}\in \R^{N_p \cdot N}$ for the vector representation of the parameters at $N=N_x \cdot N_y$ discrete 2D positions. In qMRI, one considers the forward model given by
\begin{equation}
\vec{k} = \left(\op{A} \circ q\right) \left(\vec{p}\right) + \vec{e},
\end{equation}
where $\vec{k}\in \C^{N_r\cdot  N_c \cdot N_k}$ is the vector of recorded k-space data, $\op{A}$ is a linear acquisition operator, $q$ a (non-linear) signal model, and $\vec{e}$ random noise. As the acquisition is performed $N_r$ times with different sampling parameters and changes in the MR sequence influencing the signal model, the forward model is given by $\op{A} \circ q :\,\R^{N_p\cdot N} \rightarrow \C^{N_r\cdot N_c\cdot N_k}$ with  
$\vec{p} \mapsto [ \left(\op{A}_1 \circ q_1 \right)\left(\vec{p}\right)^\trans, \ldots, \left(\op{A}_{N_r} \circ q_{N_r}\right) \left(\vec{p}\right)^\trans]^\trans$.
Here, the encoding operators at each acquisition $\op{A}_i: \,\C^{N} \rightarrow \C^{N_c \cdot N_k}$ typically are undersampled Fourier operators. These can be written as $\op{A}_i= \op{S}_i\op{F}\op{C}$
with $\op{F}$ a being two-dimensional Fourier transform and $\op{S}_i$ undersampling operators, masking out all but $N_k$ discrete points in k-space. The undersampling masks are typically varied between different acquisitions. For $N_c$ receiver coils acquiring data in parallel, the coil sensitivity operator $\op{C}$  
expands each image to $N_c$ different views acquired by the receiver coils by pixelwise multiplying with their respective spatially varying complex-valued sensitivity maps $\vec{c}_c \in \C^{N}$, often normalized such that for each pixel $\sum_i^{N_c} |c_{i}|^2 =1$. 

A qMRI reconstruction aims to obtain the tissue parameters $\vec{p}$ from the acquired data $\vec{k}$. For uncorrelated Gaussian noise $\vec{e}$, the maximum likelihood estimate is
\begin{equation}
 \vec{p^*}=\argmin_\vec{p} \|\op{A} q \left(\vec{p}\right)-\vec{k}\|^2 \,.  
\end{equation}
As the inverse problem of obtaining $\vec{p^*}$ is typically ill-posed, instead the regularized problem
\begin{equation}
\min_\vec{p} \|\op{A} q \left(\vec{p}\right)-\vec{k}\|^2 +  \mathcal{R}(\vec{p})
\label{eq:problem}
\end{equation} with a regularizer $\mathcal{R}$ has to be considered.

\subsection{Differentiable Optimization}
\label{sec:optimization}
In many computer vision tasks, data-driven approaches with differentiable layers solving an inner optimization problem within a larger neural network have been used successfully \cite{Amos2017,Agrawal2019,diffcomp}. Yet so far, in medical imaging, optimization of an inner problem is mainly only used as data-consistency layers in unrolled reconstruction networks \cite{modl} for linear problems. 

In general, to incorporate optimization problems as a layer into a larger network, which can then be trained end-to-end with gradient descent algorithms, the gradients of the solution of the inner problem with respect to the inputs must be calculated. For data-consistency layers in linear problems, this can, for example, be achieved by automatic differentiation through the steps performed by the inner optimizer if each operation is differentiable \cite{sriram}. The downsides are 1) all intermediate results have to be kept available, \rev{resulting in a linear dependence of the memory required during training on the number required iterations}, and 2) algorithms for solving non-linear problems typically contain non-differentiable operations \cite{numopt}. Alternatively, for linear problems, the gradient of the output with respect to the previous solution estimate can be efficiently calculated with matrix calculus \cite{modl,hammernikbook}. A more general approach to obtain gradients through an optimization layer uses a technique known as \textit{implicit differentiation} \cite{Gould2016}. We first revise the concept before applying it in \Fref{sec:backward}.

Let $\mathcal{F}_{\vec{\alpha}}:\R^n\rightarrow\R$ be the twice continuously differentiable objective of the inner optimization problem solved by the layer, $\vec{\alpha} \in \R^p$ the parameters to backpropagate the gradient for and $\vec{f}: \R^p \times \R^n \rightarrow \R^n$ with $\vec{f}(\vec{\alpha}, \vec{x}) =\nabla_x \mathcal{F}_{\vec{\alpha}}(\vec{x})$ be the gradient of the objective function. For some $\vec{x^*} \in \R^n$ to be a minimizer of $\mathcal{F}$, the condition
\begin{align}
    \vec{0} = \vec{f}\left(\vec{\alpha},\vec{x^*}(\vec{\alpha})\right)
    \label{eq:app_optimalcondition}
\end{align} has to be fulfilled. The well-known implicit function theorem (IFT) in a suitable notation \cite{crowder,Oliveira2013}  states:
\newtheorem{theorem}{Theorem}
\begin{theorem}[Implicit Function Theorem]
Let $\vec{f}: \R^p \times \R^n \rightarrow \R^n$ be a continuous differentiable function,
$\vec{\alpha_0} \in \R^p$ and $\vec{x_0} \in \R^n$ such that $\vec{f}(\vec{\alpha_0},\vec{x_0})=\vec{0}$ with non singular Jacobian $\pdv{\vec{f}}{\vec{x}}\left(\vec{\alpha_0},\vec{x_0}\right) \in \R^{n \times n}$. Then, there exist an open set $S \subset \R^n$ with $\vec{\alpha_0} \in S$  and a unique continuously differentiable function $\vec{x^*}: S \rightarrow \R^n$ such that $\vec{x^*}(\vec{\alpha_0})=\vec{x_0}$ and $\vec{f}(\vec{\alpha},\vec{x^*}(\vec{\alpha}))=\vec{0}$ for all $\vec{\alpha} \in S$.
\end{theorem}
Hence, \Fref{eq:app_optimalcondition} implicitly defines the minimizer $\vec{x^*}$ as a function of $\vec{\alpha}$.  Differentiating both sides of \Fref{eq:app_optimalcondition} yields
\begin{align}
   \vec{0}&=\pdv{\vec{f}(\vec{\alpha}, \vec{x^*}(\vec{\alpha}))}{\vec{\alpha}} = \pdv{\vec{f}}{\vec{x^*}}\pdv{\vec{x^*}(\vec{\alpha})}{\vec{\alpha}} + \pdv{\vec{f}}{\vec{\alpha}}  \,,
\end{align} where $\vec{x^*}$ without explicit dependency on $\vec{\alpha}$ it is treated as a fixed value. By rearranging we obtain an expression for the Jacobian of the solution mapping, i.e
\begin{align}   
   \pdv{\vec{x^*}(\vec{\alpha})}{\vec{\alpha}} &= - \left(\pdv{\vec{f}}{\vec{x^*}}\right)^{-1}\pdv{\vec{f}}{\vec{\alpha}} \,.
\end{align} 
Therefore, by the chain rule, given the gradient of an outer loss $\pdv{\mathcal{L}}{\vec{x}}$ as a row vector at $\vec{x}=\vec{x^*}$, the row vector $\pdv{L}{\vec{\alpha}}$ at $\vec{\alpha}=\vec{\alpha_0}$ can be written as 
\begin{align}
 &\pdv{\mathcal{L}}{\vec{\alpha}}\/(\vec{\alpha_0})=\pdv{\mathcal{L}}{\vec{x}} \/(\vec{x^*})\pdv{\vec{x^*}(\vec{\alpha})}{\vec{\alpha}}\/(\vec{\alpha_0},\vec{x^*}) \\
   &= - \pdv{\mathcal{L}}{\vec{x}} \/(\vec{x^*}) \left(\pdv{\mathcal{F}}{\vec{x}}{ \vec{x}^\trans}\/(\vec{\alpha_0},\vec{x^*})\right)^{-1}\pdv{\mathcal{F}}{\vec{\alpha}}{\vec{x}}\/(\vec{\alpha_0},\vec{x^*}) \,.
\end{align} 
Finally, dropping the explicit points of evaluation in the notation, we obtain the column vector 
\begin{align}
  \left(\pdv{\mathcal{L}}{\vec{\alpha}}\right)^\trans &= - \pdv{\mathcal{F}}{\vec{\alpha}}{\vec{x}} 
  \left(\pdv{\mathcal{F}}{\vec{x}}{ \vec{x}^\trans}\right)^{-1} \left(\pdv{L}{\vec{x}}\right)^\trans \,,
  \label{eq:implicitdiff}
\end{align} where we have used the symmetry of the Hessian $\pdv{\mathcal{F}} {\vec{x}} {\vec{x}^\trans}$. 

This gives the general blueprint to implement the backpropagation step for optimization layers in deep-learning frameworks. 
Either the analytical Hessian or the vector-Hessian-product functionality of the framework can be used in conjunction with an iterative solver for linear problems to approximately obtain $(\pdv{\mathcal{F}}{\vec{x}}{ \vec{x}^\trans})^{-1} \left(\pdv{\mathcal{L}}{\vec{x}}\right)^\trans$ in \Fref{eq:implicitdiff}. For an approximate solution with an error with norm $\epsilon$, it has been shown that the error of the estimated gradient is $\mathcal{O}(\epsilon)$ \cite{pedregosa}. The derivative with respect to the parameters of the gradient of the inner optimization loss used in \Fref{eq:implicitdiff}, $\pdv{\mathcal{F}}{\vec{\alpha}}{\vec{x}}$, can also either be calculated analytically or by autograd functionality.

\subsection{Proposed Reconstruction Network PINQI}
For solving the qMRI inverse problem, we propose the following, physics-informed end-to-end approach \textit{PINQI}:
  First, we choose the regularization in \Fref{eq:problem} such that the objective is finding
\begin{equation}
\vec{p}^*=\argmin_{\vec{p}}
  \| \op{A} q(\vec{p}) - \vec{k} \|_2^2 + \lambda_{\tilde{p}} \| \vec{p} - \vec{p}_{\mathrm{reg}}\|_2^2 \,,
\label{eq:reg_problem}
\end{equation} with $\vec{p}_{\mathrm{reg}}$ a regularizing prediction for the parameters. We introduce an auxiliary variable $\vec{y}:=q(\vec{p})$ and include equality by a quadratic penalty constraint \rev{as well as an additional regularizing prediction for the images $\vec{y}_{\mathrm{reg}}$}, 
\begin{align}
\label{eq:minyp}
    \min_{\vec{y},\vec{p}} &\|\op{A} \vec{y}-\vec{k}\|_2^2+\lambda_{\tilde{p}}\left\|\vec{p}-\vec{p}_{\mathrm{reg}}\right\|_2^2\\&+\lambda_q\|q(\vec{p})-\vec{y}\|_2^2 + \lambda_y \| \vec{y} - \vec{y}_{\mathrm{reg}}\|_2^2  \nonumber \,,
\end{align}
where all $\lambda$ are positive regularization strengths. Next, we split Problem \eqref{eq:minyp} into two subproblems which we solve in an alternating fashion \cite{Afonso2010} within our unrolled reconstruction network.

\begin{figure*}
\centering
\includegraphics[width=0.75\linewidth]{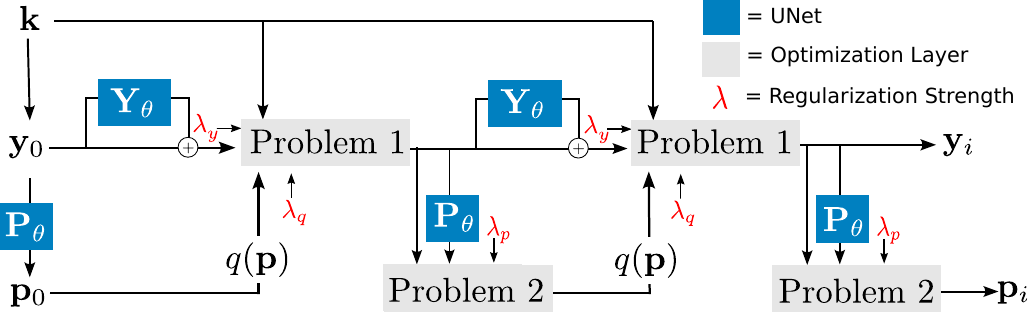}
\caption{Schematic of the unrolled physics-informed network to solve \Fref{eq:reg_problem} by quadratic splitting as used by our proposed PINQI. We alternate between solving two subproblems, Problem 1 is a linear data-consistency problem and solved by a differentiable conjugate-gradient block. Subproblem 2 is solved by a differentiable non-linear optimization block. $\op{Y}_\theta$ denotes a residual UNet operating on qualitative images, $\op{P}_\theta$ the parameter prediction UNet. The predictions of these subnetworks are used as regularizers with (learnable) strength $\lambda_y$ and $\lambda_p$, respectively. Consistency between both subproblems is relaxed to a quadratic penalty weighted by $\lambda_q$.  For more details regarding the formulations of the two subproblems, see the main text. }
\label{fig:network}
\end{figure*}
\subsubsection{Linear Subproblem: Optimization in $\mathbf{y}$} The subproblem of obtaining artifact-free qualitative images 
\begin{align} 
\label{eq:sub_linear_F}
\vec{y}^*=
\argmin_{\vec{y}}\|\op{A} \vec{y}-\vec{k}\|_2^2 &+\lambda_y\|\vec{y}-\vec{y}_\mathrm{reg}\|_2^2 \\&+\lambda_q\|q(\vec{p})-\vec{y}\|_2^2 \nonumber
\end{align}
is similar to the problem commonly solved in MR image reconstruction networks \cite{modl} by the data-consistency blocks, but extended by the model-based  \cite{Block2007,tran2013,becker2019} last term which penalizes a discrepancy between reconstructed qualitative images and predictions based on the estimate for the quantitative parameters.  As we consider multi-coil imaging, the minimizer of \Fref{eq:sub_linear_F} does not have a closed-form analytical solution. Instead, we solve a problem of the form $\op{H}{\vec{x}}=\vec{b}$ with $\op{H}=\op{A}^{\herm} \op{A} + (\lambda_q  + \lambda_y)\, \unity_{N}$ and $\vec{b}=\op{A}^\herm \vec{k} +\lambda_q\, q(\vec{p}) + \lambda_y\, \vec{y}_\mathrm{reg}$ approximately, for example
with the conjugate gradient (CG) algorithm \cite{numopt}.

\subsubsection{Non-Linear Subproblem: Optimization in $\mathbf{p}$}
 By introducing $\lambda_p=\sfrac{\lambda_{\tilde{p}}}{\lambda_q}$, the second subproblem can be written as finding 
\begin{equation}
   \vec{p^*}= 
\argmin_{\vec{p}}\|q(\vec{p})-\vec{y}\|_2^2 +\lambda_p\left\|\vec{p}-\vec{p}_{\mathrm{reg}}\right\|_2^2\,.
    \label{eq:sub_non-linear_F}
\end{equation}
Due to the non-linear signal function $q$, this subproblem is non-linear. Hence, we introduce the non-linear optimization layer depicted in \Fref{fig:nonlinear_layer}. This layer uses L-BFGS \cite{bfgs,numopt} to approximately solve the problem in the forward pass. Depending on the concrete signal model, within this layer, different solvers and preconditioning techniques might be chosen \cite{practical} instead.

Finally, we construct our proposed PINQI as shown in \Fref{fig:network} by alternating between both subproblems for a fixed number of iterations. In each iteration $i=1\ldots T$, we obtain predictions for the qualitative, artifact-free images $\vec{y}^i$ and for the quantitative parameters $\vec{p}^i$. For regularization, we use predictions obtained by trained subnetworks with parameters $\vec{\theta} \in \R^n$ as $\vec{y}^i_\mathrm{reg}=\vec{y}^i_{\theta}=\op{Y}_\theta(\vec{y}^{i-1},i)$ and $\vec{p}^i_\mathrm{reg}=\vec{p}^i_{\theta}=\op{P}_\theta(\vec{y}^{i},i)$, respectively.  All learnable parameters of the UNet subnetworks are shared between iterations.

\subsection{ Gradients of the Subproblem Solutions}
\label{sec:backward}
We propose to train PINQI end-to-end, i.e.\ we construct an objective function $\mathcal{L}$ which depends on the final predicted real-valued quantitative parameter maps. Also, we aim to use gradient descent-based algorithms to optimize the learnable network weights $\vec{\theta}$ and use backpropagation to find the direction of the steepest descent. Thus, we need to be able to calculate the gradients of the solutions found by the solvers of both subproblems with respect to all variables depending on $\vec{\theta}$. We achieve this by implementing the solvers as differentiable optimization layers as presented in \Fref{sec:optimization}.

\begin{figure}
    \centering
    \includegraphics[width=\onecol]{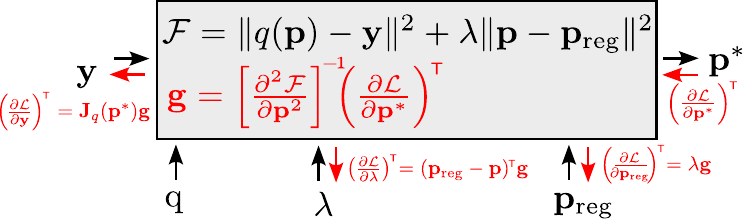}
    \caption{Proposed non-linear optimization layer, finding ${\vec{p}^*=\argmin_\vec{p} \mathcal{F}}(\vec{p})$ with an off-the-shelf solver while allowing backpropagation of the gradients (red) to the regularization parameters, $\lambda$ and $\vec{p}_\mathrm{reg}$, and data $\vec{y}$.}
    \label{fig:nonlinear_layer}
\end{figure}

First, we specialize the general concept to a regularized non-linear regression problem, i.e.\ an inner problem of obtaining $\vec{p^*}{=} \argmin_{\vec{p}} \mathcal{F}(\vec{p})$
with an objective 
\begin{equation}
    \mathcal{F}(\vec{p})=\|q(\vec{p})-\vec{y}\|_2^2 + \lambda \|\vec{p}-\vec{p}_\mathrm{reg}\|^2 \,,
    \label{eq:inner}
\end{equation} 
 as used in subproblem 2. Suppose the gradient of the inner objective at $\vec{x}^*$ is continuously differentiable and its Jacobian is invertible. Then, the required gradient of $\mathcal{L}$ with respect to the trainable weights can simply be found by applying \Fref{eq:implicitdiff}. The resulting equations for the propagation of gradients to $\lambda$, $\vec{p}_\mathrm{reg}$ and $\vec{y}$ are shown in \Fref{fig:nonlinear_layer}, which summarizes our proposed non-linear regression layer for this subproblem. Within this layer, we use CG to approximate $\vec{g}$, which denotes the product of the inverse Hessian of $\mathcal{F}$ with the gradient of $\mathcal{L}$ with respect to the output of the solver. We employ automatic differentiation to 
 compute the Hessian-vector product applied inside the CG algorithm, as well as to compute the  derivative (with respect to $\vec{\theta}$) of the gradient (with respect to $\vec{p}$) of the inner loss. 
In \Fref{sec:regressionexp} we investigate the influence of including a single instance of our differentiable non-linear optimization layer at the end of an otherwise unchanged qualitative image reconstruction network. 

Similarly, backpropagation through the solver of the linear image reconstruction subproblem 1 can also be efficiently calculated, as demonstrated in \cite{modl} for gradients with respect to $\vec{y}_{\mathrm{reg}}$. Extending upon these results, we use implicit differentiation to derive the gradients with respect to all inputs:\ For an inner optimization objective of the form
\begin{equation}
    \mathcal{F}(\Vec{y})= \|\op{A}\vec{y} - \vec{k}\|^2 +\sum_i\lambda_i \|\vec{y} - \vec{y}_\mathrm{i}\|^2 \,,
\end{equation} 
and a solution $\vec{y^*} \in \C^N$, that fulfills the necessary  condition 
$\nabla F(\vec{y^*})=0$, application of \Fref{eq:implicitdiff} and defining 
\begin{equation}
    \vec{g} := (\op{A}^\herm\op{A} + \textstyle \sum_i \lambda_i \unity_N )^{-1} \left(\pdv{\mathcal{L}(\vec{y}^*)}{\vec{y}^{*H}}\right)^\trans
\end{equation} 
gives the gradients of the outer loss (in Wirtinger calculus \cite{wirtinger}): 
\begin{align}
\pdv{\mathcal{L}}{\vec{y}_\mathrm{i}^\herm} =  \lambda_i \vec{g}\,, &&
\pdv{\mathcal{L}}{\lambda_i} = \Re{(\vec{y}_i-\vec{y})^\herm \vec{g}}\,,  &&
\pdv{\mathcal{L}}{\vec{k}^\herm}= \op{A} \vec{g}\,. 
\end{align}
For multi-coil MRI, the linear operator defining the data recorded in a single acquisition by the $j$-th coil is of the form $\op{A}_j= \op{S}\op{F} \op{C}_j$, and,
\begin{align}
\pdv{\mathcal{L}}{\vec{c}_j^\herm} &= \left(\op{F}^\herm \left(\op{A}_j\vec{y^*}-\vec{k}_j\right) \right) \odot\overline{\vec{g}} + \left(\op{F}^\herm \op{A}_j \vec{g}\right) \odot \overline{\vec{y}}^*
\label{eq:csmgrad}
\end{align}
can be used to backpropagate the gradient of the outer loss to the $j$-th estimated unnormalized sensitivity map $\vec{c}_j$. Here $\odot$ denotes the Hadamard product, $\overline{\vec{g}}$ the complex conjugate of $\vec{g}$, and $\vec{k}_j$ the data recorded by the $j$-th coil.

\section{Application to $T_1$-Mapping}

The proposed PINQI can be used for many quantitative MR imaging techniques by adapting $q$ to the respective signal model and $\op{A}$ to the acquisition operator.\\
For our experiments, we chose a $T_1$-mapping of the brain using a saturation recovery sequence as a typical, yet conceptually simple, qMRI problem.  Here,  the signal model $q_i$ at the $i$-th acquisition is given for each pixel by
\begin{equation}
q_i(\Re\,M_0, \Im\,M_0, T_1) = M_0 (1-\exp(-\tau_i/T_1))
\end{equation}
with $\mathrm{Re}\,M_0$ and $\mathrm{Im}\,M_0$ denoting the real and imaginary part of the complex initial magnetization $M_0$, $T_1$ the longitudinal relaxation time, and $\tau_i$ the $i$-th saturation recovery time, i.e.\ the delay between the magnetization preparation pulse and the data acquisition.
As encoding operator $\op{A}$, we chose a Fourier transform with Cartesian sampling, $4$-$8$-fold undersampling, and 8 receiver coils. 

All code relating to \rev{a PyTorch implementation of} PINQI, \rev{the optimization layer}, our implementations of the reference methods, the synthetic data generation as well as the MR sequences will be made available after peer-review at \href{https://github.com/fzimmermann89/PINQI}{github.com/fzimmermann89/PINQI}.
\subsection{Training Dataset and Data Aquisition}
We utilized synthetic data for training and validation and conducted testing on both synthetic data and real scanner data. 
\subsubsection{Synthetic Data}
The synthetic data was generated from the BrainWeb dataset \cite{brainweb}, which consists of three-dimensional segmentation masks of 20 healthy human brains, which we split into 16/1/3 for training/validation/testing. During the training process, we randomly assigned anatomically plausible values\cite{bojorquez2017} for $M_0$ and $T_1$ to each of the 11 tissue classes in every sample on-the-fly. These values were combined with axial slices of the masks, resampled to $192 \times 192$ pixels, into initial synthetic parameter maps. To increase spatial variability and prevent overfitting to piecewise constant images, we augmented the $T_1$-maps by multiplying them with low-variance 2D polynomials. Similarly, we augmented the $M_0$-maps by a random spatially slowly varying complex phase and bias field. As additional augmentations during training, we performed flips and rotations $<$10°. The resulting parameter maps were considered ground truth labels.
In addition, we generated sample-specific masks indicating the presence of brain tissue within the maps. During training, these masks were used to weigh down the losses outside the brain region. During the testing phase, all quantitative measures used to report the performance of the methods were restricted to the relevant brain tissue. 

\rev{The saturation recovery times were set as either 0.5\,s, 1\,s, 1.5\,s, 2\,s, and 8\,s (synthetic comparison, phantom, and in-vivo experiments),  or as 0.5\,s, 0.7\,s, 0.9\,s, 1.1\,s, 1.3\,s, 1.6\,s, 2\,s, and 8\,s  (ablation study).}  We used variable density 1D-under-sampling with 8 \rev{randomly generated \cite{torchkbnufft} bird-cage-like receiver coils with random orientation. For each coil, we used a Gaussian amplitude profile with randomly varying half-width in $[0.2, 0.5]$ times the field-of-view and a random phase modulated by a slowly varying random 2D polynomial}. The undersampling patterns were randomly chosen for each recovery time and for each sample. Finally, for each sample, we added complex Gaussian noise with  randomly chosen standard variation $\sigma \in (0.001,0.04)$ to simulate noisy measurements. 

Our synthetic validation and testing datasets consisted of a fixed set of generated labels and simulated noisy undersampled k-space measurements. \rev{Here we used the ground truth sensitivity maps, thus assume a fully known forward model.}

\subsubsection{Data Aquisition}
For further validation of our proposed method, we acquired two sets of scanner data on a $3$\,T MRI system (Siemens Verio): 1) data from a physical phantom and 2) data from the brains of healthy volunteers. The examination was approved by the local ethics committee and is in accordance with the relevant guidelines and regulations. Written informed consent was received from all subjects prior to the examination. 
The phantom consisted of 9 tubes filled with liquids prepared to have $T_1$ times in the range of approx. $250$\,ms-$1750$\,ms.  
We used adiabatic saturation pulses, 5 saturation recovery times, spoiled GRE readouts with inside-out order and $1$\,mm$\times1$\,mm$\times8$\,mm resolution, $192\times192$ matrix size, $6\deg$ flip angle, a 32-channel head coil. For each saturation time, the 1D Cartesian undersampling was chosen randomly, analogously to our training. As a reference, we performed fully sampled saturation recovery measurements with $18$ delay times and obtained the $T_1$ values by pixel-wise regression on the reconstructed images. Coil sensitivity maps were either estimated from the \rev{autocalibration region, i.e.\ the fully sampled 12/10/8 central lines, of the 4x/6x/8x undersampled} image with the longest $\tau$, or, for the reference, from the fully-sampled image\cite{walsh,inati,espirit}. All MR sequences were implemented using the vendor-agnostic Pulseq framework \cite{pulseq}.
\subsection{PINQI: Implementation Details}

We set the number of alternations between the subproblems in our implementation of PINQI as $T=5$ (empirically chosen). In the non-linear subproblem, we used the L-BFGS algorithm\cite{bfgs,numopt} with a trigonometric transformation of the variables \cite{practical} to  enforce the bounds $\Re(M_0) \in (-2,2)$, $\Im(M_0) \in (-2,2)$, and $R_1:=1/T_1 \in (-1 s^{-1}, 20 s^{-1})$. These bounds are much wider than any plausible predictions and only served to increase stability at the beginning of training. \rev{We initialized the solver in the first iteration with $\vec{p}_{\mathrm{reg}}$, in subsequent iterations with the result obtained in the previous iteration.}
\rev{The linear problems in \Fref{eq:sub_linear_F} and \Fref{eq:implicitdiff} were approximately solved by CG\cite{numopt} with the norm of the residual $<10^{-6}$ as stopping criterion.}
Both regularizing networks, $\op{P}_\vec{\theta}$ and $\op{Y}_\vec{\theta}$, are UNets \cite{ronneberger2015} with residual blocks, each consisting of two SiLU\cite{silu} activated convolutions (window size 3) and group normalizations (group size 16)\cite{groupnorm}. In 
$\op{Y}_\vec{\theta}$, the two convolutions of each block handle all temporal points as batched samples, and a third convolution operates along the temporal direction, handling all spatial points as batched samples \cite{Qiu2017}.
In both UNets, we condition on the iteration of the unrolled reconstruction by performing learned projections to scale and shift values for each feature map \cite{film,nosense} after the first convolution in each encoder and decoder block. The downscaling in the UNet is done with stride 2, kernel 2 max-pooling operations in the spatial dimensions. Each upscaling is performed with 2x bilinear interpolation in the spatial dimensions followed by a 3x3 convolution. The number of output features at the different layers in $\op{Y}_\vec{\theta}$ is empirically chosen as  (16, 32, 48, 64), and as (32, 64, 96, 128) in $\op{P}_\vec{\theta}$. This results in 597,323 and 2,235,308 learnable weights, respectively.  We found initializing the network such that each block mainly operates pixel-wise \cite{rezero} improved training stability.

All regularization strengths $\lambda_{p,i}$, $\lambda_{y,i}$, and $\lambda_{q,i}$ are iteration-dependent learnable parameters. We enforced the positivity of these parameters by setting $\lambda_{i}=\log \frac{1}{5}(1+\exp\{5\tilde{\lambda_i}\})$ and optimizing for each $\tilde{\lambda}_i$. We empirically chose an initialization corresponding to $\lambda_{p,i} = 3$, $\lambda_{y,i} = 0.1$, and $\lambda_{q,i}=0.1+0.05i$.
The training was performed by minimizing the MSE between the predicted quantitative parameters, $R_1=1/T_1$, $\Re(M_0), \Im(M_0)$, and the corresponding target parameters. Deep supervision \cite{dudornet,deepsupervision} was applied by also incorporating the MSE loss of the predicted parameters during all previous iterations, strongly weighted down by a factor of $0.05$.
We pretrained $\op{P}_\vec{\theta}$ for one epoch with random linear interpolation between the zero-filled reconstructions of the noisy k-space data and the ground truth (noise- and artifact-free) qualitative images as input with MSE loss on the estimated parameters. The training was performed for 80 epochs. Both training phases were performed using the Adam optimizer \cite{adam} with a maximum learning rate $0.004$ for the UNet parameters and $0.001$ for all $\bar{\lambda}$'s, a linear warmup and cosine learning rate schedule \cite{he2019}, weight decay \cite{adamw} $0.01$, and a batch size of 16.

\begin{figure*}
    \raisebox{2.7cm}{\rotatebox{90}{ \small $T_1$\hspace{5cm}$|M_0|$}}
    \centering
    \newcommand{\PlotSynth}[1]{%
    \begin{tikzpicture}
        \node[anchor=south west] (image) at (0,0) {
             \includegraphics[width=0.135\textwidth]{#1}
        };
    \end{tikzpicture}
    }%
    \PlotSynth{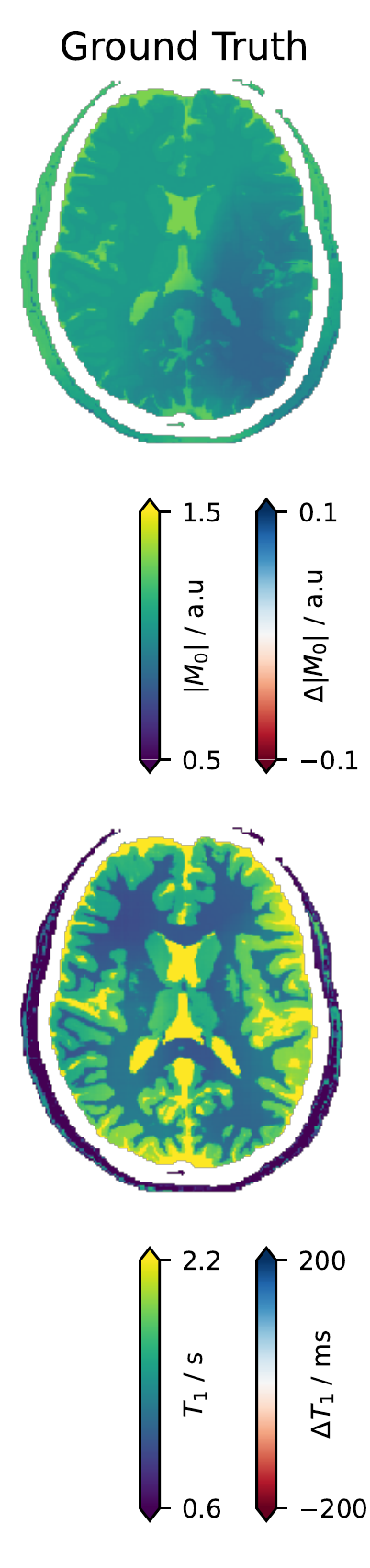}
    \PlotSynth{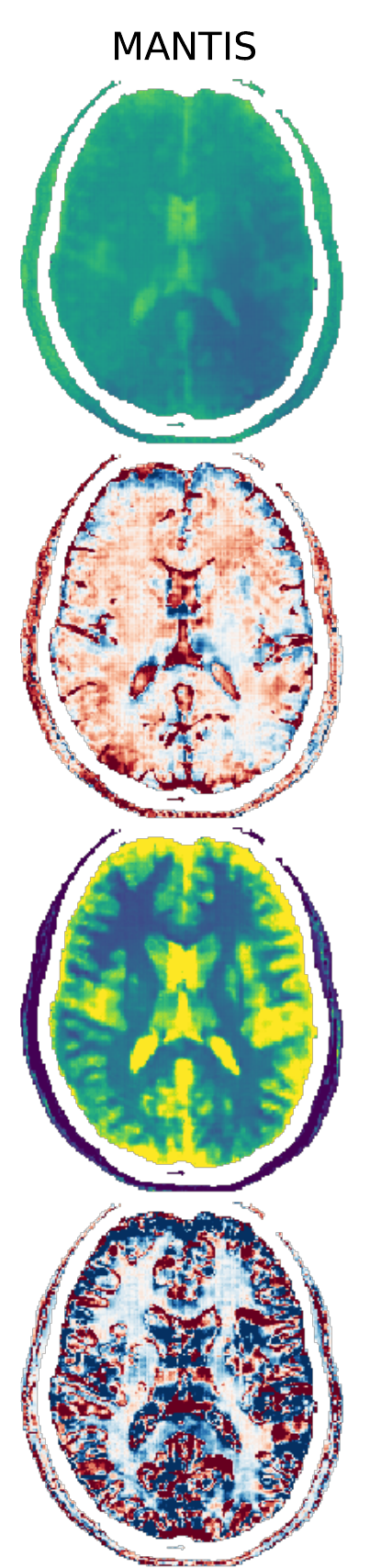}
    \PlotSynth{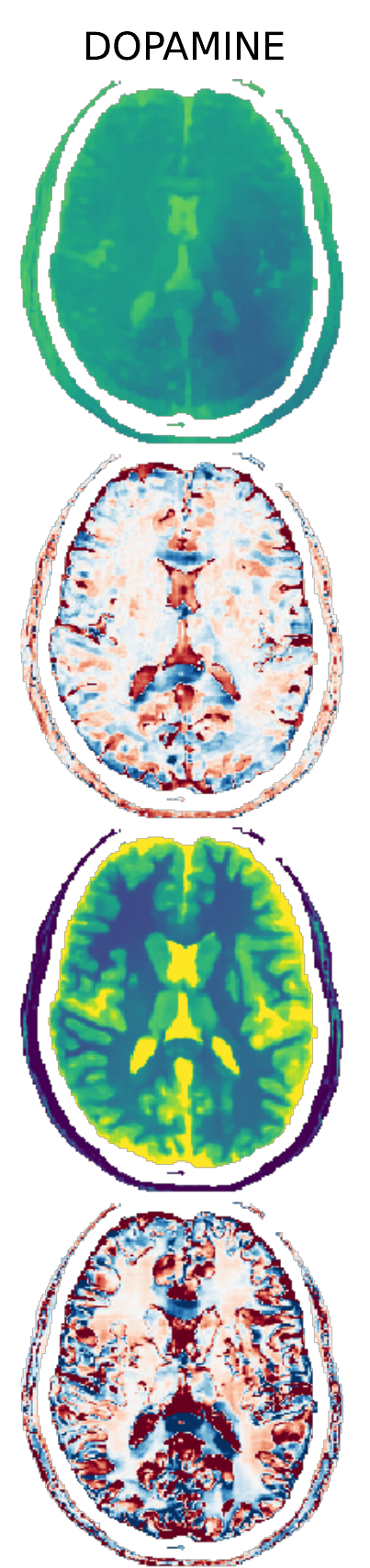}
    \PlotSynth{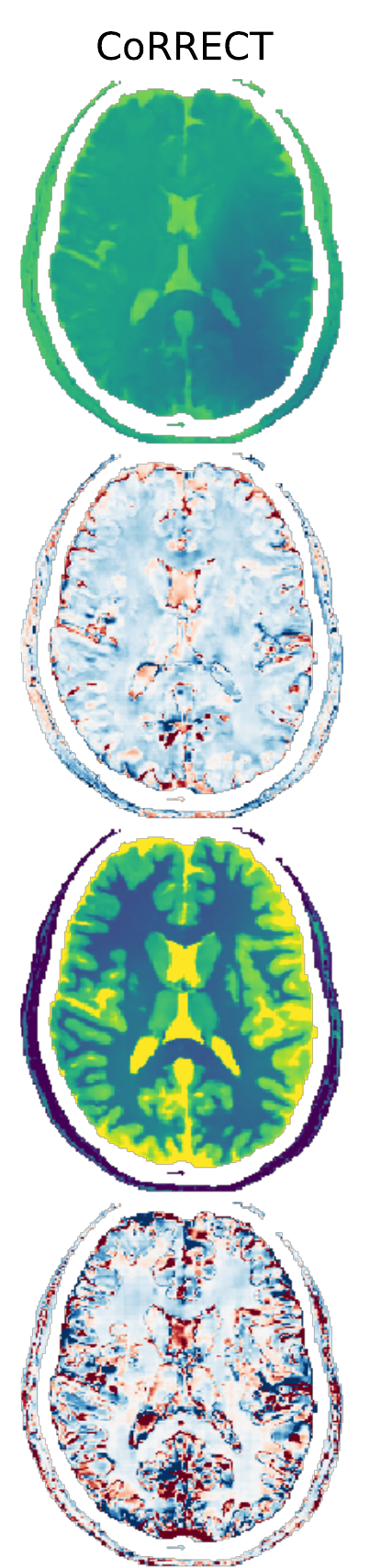}
    \PlotSynth{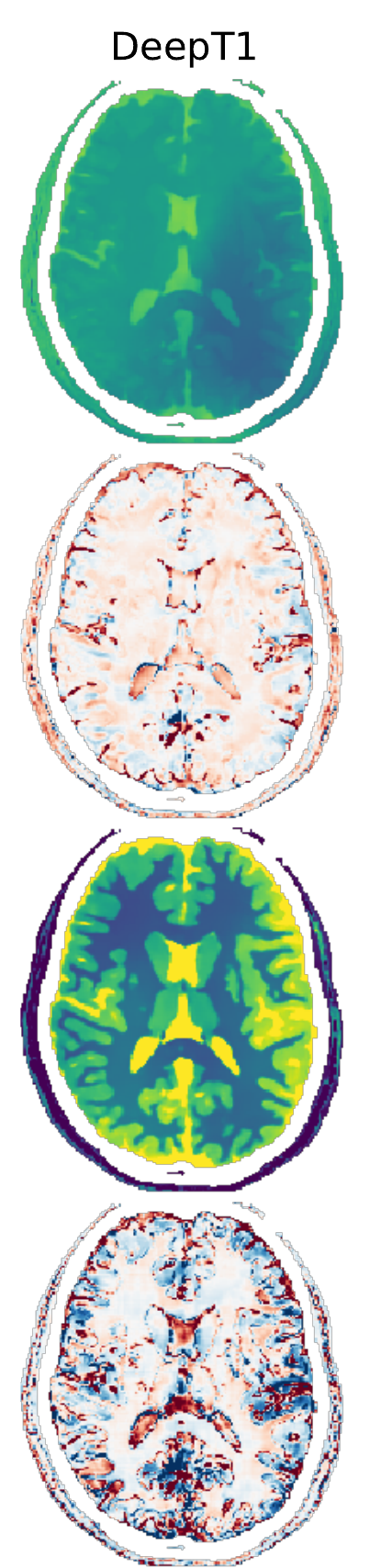}
    \PlotSynth{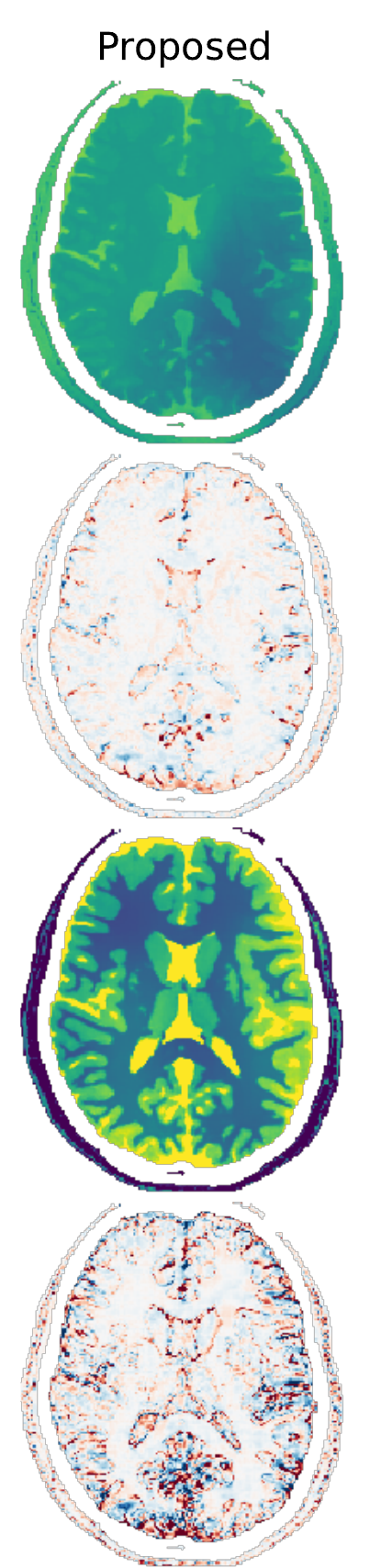}

    \caption{Examplary results of the different methods for one simulated measurement from the test dataset \rev{at 8-fold acceleration}. More details regarding the different methods are provided in the text. For each method, the magnitude of $M_0$ and deviation from the ground truth are shown in the top two rows. The calculated $T_1$ and deviation from ground truth are shown in the bottom two rows. 
    }
    \label{fig:result_silico}
\end{figure*}
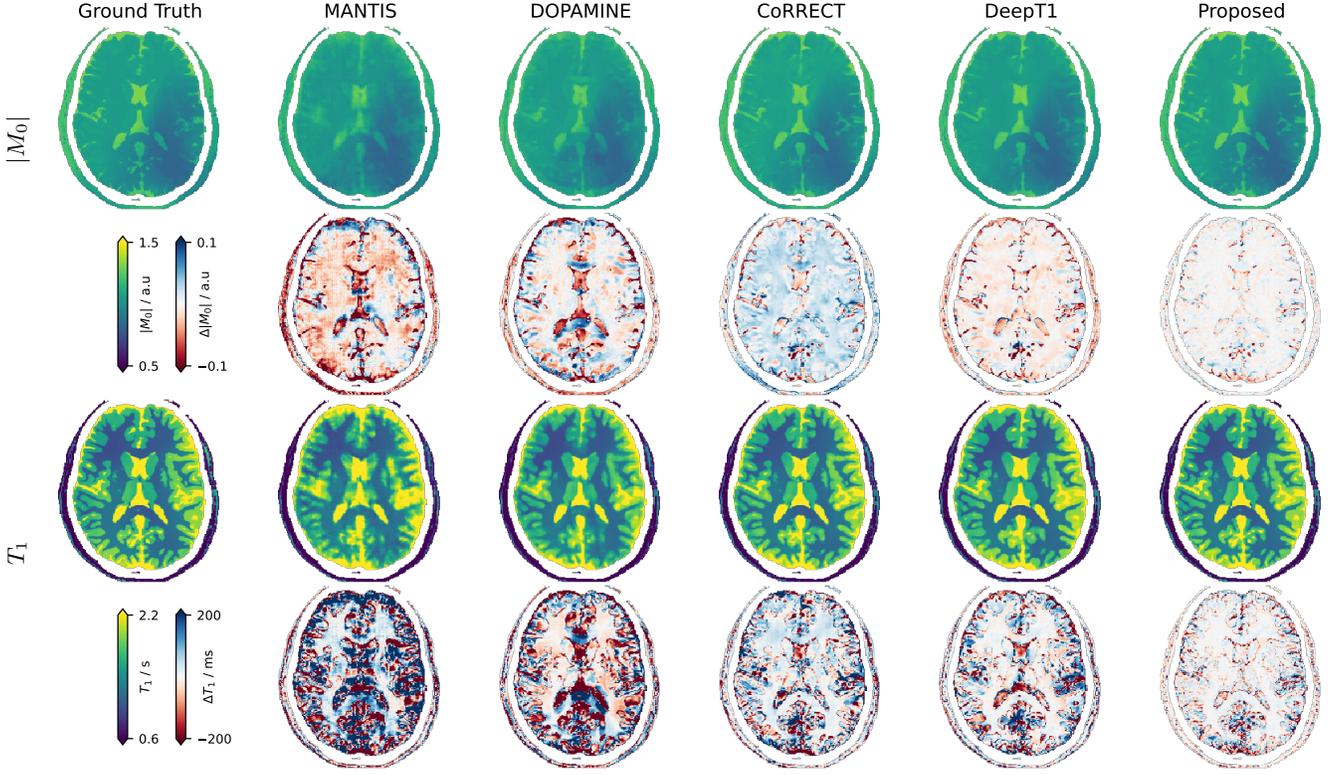

\subsection{Methods of Comparison}
Besides our proposed PINQI approach, we provide results of our implementations of \rev{four} recently published learned reconstruction methods trained and tested on the same dataset. As the general superiority of the selected learned comparison methods to non-learned methods has been established by the respective authors \cite{deept1, correct, dopamine, mantis, modl}, we abstain from comparing against non-learned baselines.

\paragraph{DeepT1 \cite{deept1}}
Our reimplementation of DeepT1 tries to stay as close as possible to the provided description in the original publication while adapting to our synthetic data and the saturation recovery signal model. The task of magnitude-only image reconstruction by an unrolled reconstruction uses a recurrent CNN (297,538 learnable weights). We replaced the original data-consistency layers, which were only suitable for a single coil setting, with CG-based ones. The parameter estimation from the reconstructed magnitude images was done with a UNet (20,547,906 learnable weights). 
\rev{
\paragraph{CoRRECT \cite{correct}}
We implemented the network of CoRRECT as described in the original publication, using a CNN (187,010 learnable parameters) in the unrolled image reconstruction and a UNet (46,741,698 parameters) for the parameter estimation. We adapted the method to our setting. 
Besides a supervised reconstruction loss and a self-supervised parameter estimation loss, we also included a supervised MSE loss of the quantitative parameters. This addition was motivated by the availability of ground truth labels in our setting and improved the results on our validation set, ensuring a fair comparison. For combining the loss terms, we empirically found weighting factors of $1$, $0.1$, and $1$, respectively.
}
\paragraph{MANTIS \cite{mantis}}
We adopted the MANTIS method for $T_1$-mapping and training on our simulated dataset. As MANTIS takes zero-filled \textit{magnitude} images as input, it cannot predict the phase of $M_0$. The particular UNet architecture proposed for MANTIS uses 29,248,258 learnable parameters. We used the proposed combination of the MSE of the quantitative parameters and in k-space as loss and tried different weighting of the latter. Although we noticed only a minimal influence on the performance on the validation set, we report results for a network trained with the optimal weighting found in our experiments of $0.01$.
\paragraph{DOPAMINE \cite{dopamine}}
We modified the DOPAMINE method for our signal model and supervised training. Our implementation performs 10 iterations of unrolled gradient descent. Each overall stepsize and each weighting of the neural network predicted step were scalar trainable parameters with softplus enforced positivity. The two regularizing CNNs and the network for starting point prediction combined had 317,894 trainable parameters in total. The training was performed on MSE of $\vec{p}$. As the training was highly unstable, we slightly modified DOPAMINE by soft-clipping each update in the unrolled gradient descent via a scaled $\tanh$. A clipping threshold higher than the parameters' dynamic range over the training data was sufficient to stabilize training. We emphasize that during inference on the validation set, each update step proposed by the network was much smaller than allowed by the clipping, thus our modification should not negatively influence test performance.

\subsection{Ablation Study}
\label{sec:ablation}
\label{sec:regressionexp}
Ablations of different parts of our proposed unrolled network are used to demonstrate the incremental benefits of the major components of our method. For each ablation, the respective learning rate of the parameters of the UNet was tuned based on the validation set MSE. The study was performed for eight acquisition times \rev{and 8-fold undersampling}.

\paragraph{\rev{No Signal Function}}
\rev{To highlight the benefit of incorporating the knowledge about the signal function into the network through the proposed unrolling, we instead} consider, similar to DeepT1 \rev{and CoRRECT}, solving the two problems of image reconstruction and parameter estimation only once.
We use an unrolled model-based image reconstruction with a UNet-predicted regularization (5 iterations of network application and data-consistency) \cite{modl,sriram}. The quantitative parameters are predicted by a separate UNet. Both UNets have the same architecture as the corresponding ones in PINQI. We performed a pretraining of the reconstruction network for 5 epochs (optimizing on MSE of the complex-valued qualitative images) followed by end-to-end optimization for 60 epochs. This baseline approach does not use the non-linear signal model $q$ at inference time but benefits from architectural improvements to the CNNs over the comparison methods.
\rev{
\paragraph{No Image-Space 
Regularization}
We set $\lambda_y = 0$ and remove $\op{Y}_\theta$ from PINQI to investigate the importance of the learned regularization in image-space while retaining the regularization in the non-linear subproblem. The training setup and all other parameters remain the same.
\paragraph{No Parameter-Space Regularization} We set $\lambda_p = 0$ and remove $\op{P}_\theta$ from PINQI, effectively removing the learned regularization in parameter-space. Instead, we only perform a non-regularized regression on the signal model in the non-linear subproblem 2. All other parameters remain unchanged.
\paragraph{No Non-Linear Solver}
We remove the proposed differentiable non-linear optimization layer for subproblem 2 from our network and instead directly consider the network prediction of the parameters as the solution of \Fref{eq:sub_non-linear_F}. Note that we still iterate between the subproblems and train end-to-end. This ablation results in a learned reconstruction scheme similar to, for example, PGD-Net\cite{deepunrolling}.
\paragraph{Gradient Descent}
Instead of the proposed L-BFGS-based optimization with implicit gradient calculation, we perform 5 steps of gradient descent on \Fref{eq:sub_non-linear_F} in each alternation. We use standard backward-mode autograd to obtain the gradients of the steps and make the stepsizes for each alternation trainable parameters. The memory consumption during training is approximately 10\% higher than in PINQI.
\paragraph{Fixed $\vec{p}_\mathrm{reg}$ and $\vec{y}_\mathrm{reg}$ across all iterations}
We only apply the image-space and parameter-space UNets once, i.e.\ $\vec{p}_\mathrm{reg} =  \op{P}_\theta(\vec{y}_0)$ and $\vec{y}_\mathrm{reg} =  \op{Y}_\theta(\vec{y}_0)$, instead of updating the UNet predictions in each iteration of the unrolled reconstruction. While the number of training parameters stays constant, this significantly reduces the computation necessary in a single forward pass. We keep the number of iterations $T=5$, iterating between both subproblems to solve \Fref{eq:minyp}.
\paragraph{Single Iteration / Two Iterations}
We set either $T=1$ or $T=2$, performing either a single iteration or two iterations of PINQI, respectively. Note, that $T=1$ differs from the experiment in \textit{f)}, where we still alternate between solving two subproblems.
}

For further investigation of the advantage of incorporating differentiable non-linear regression into the reconstruction, we used the same UNet architecture as used in PINQI for $\op{Y}_\theta$ also in an unrolled reconstruction of the qualitative images $\vec{y}$ and compare the following two cases:

\paragraph{Neural Network Reconstruction and Separate Regression}
We optimized the network by minimizing the MSE of the qualitative images. At inference time, we used the resulting reconstructed images to perform a non-linear least-squares regression using BFGS to obtain $\vec{p}$. 
\paragraph{Neural Network with Non-linear Optimization Layer for End-to-End Regression}
We include the non-linear regression by means of the proposed optimization layer (\Fref{fig:nonlinear_layer} in the network. As a result, we were able to train the network in an end-to-end fashion, minimizing the MSE of the quantitative parameters. 

\section{Results}
\subsection{Comparison with Reference Methods}
\label{sec:result_synth}

\begin{figure*}
    \centering
    \includegraphics[width=1.0\onecol,trim={2cm 1.5cm 4cm 1.5cm},clip]{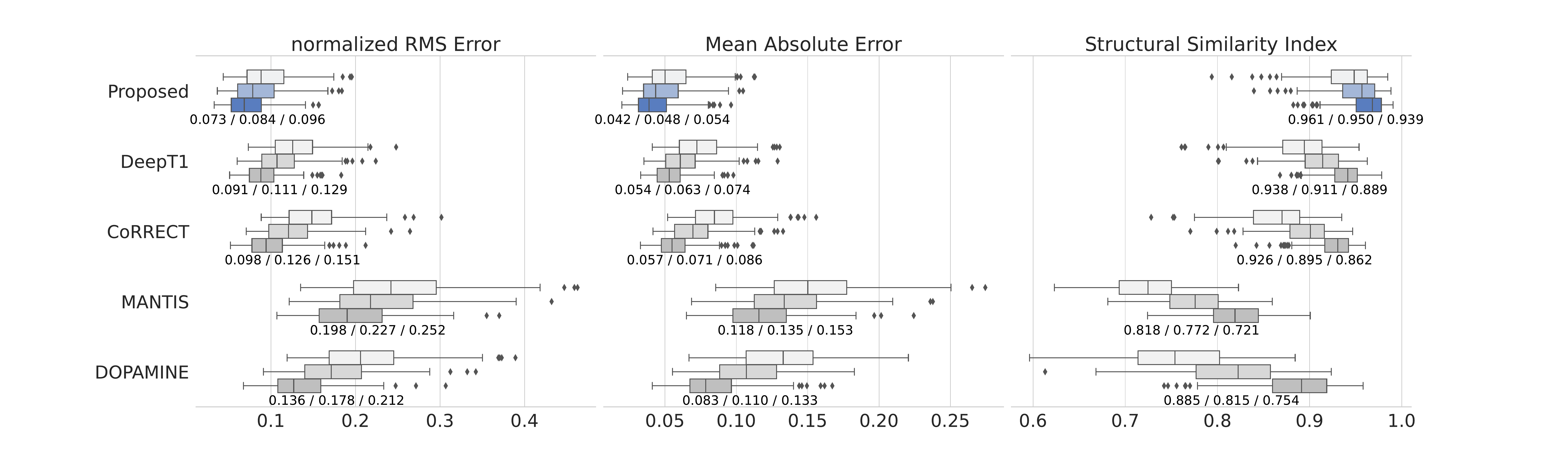}
    \caption{Comparison of our proposed PINQI with \rev{four} different state-of-the-art learned qMRI reconstruction methods \cite{mantis,deept1,correct,dopamine} in terms of nRMSE, MAE, \rev{and SSIM} of $T_1$ for each sample of the test set \rev{at 4-fold (darkest, bottom), 6-fold and 8-fold (brightest, top) undersampling}. The mean values over all samples \rev{ at 4-fold/6-fold/8-fold undersampling are provided as labels.}
    PINQI improves upon all of the comparison methods in \rev{all three} metrics.}
    \label{fig:result_silico_box}
\end{figure*}

The results of training PINQI and the comparison methods on the synthetic dataset and application on the test set are shown in \Fref{fig:result_silico_box}. Our proposed PINQI method achieves \rev{at the highest acceleration factor we investigated, 8-fold,} a normalized root mean squared error (nRMSE) of the $T_1$ maps of less than $0.10$ whereas the lowest nRMSE achieved by one of the comparison methods is $0.13$ (DeepT1 and the similar performing CoRRECT) and both MANTIS and DOPAMINE have nRMSE exceeding $0.2$. The mean absolute error (MAE) of $T_1$ is again lowest for PINQI ($0.05$) compared to $0.07$, $0.15$, and $0.13$ for DeepT1, MANTIS, and DOPAMINE, respectively. \rev{In terms of the Structural Similarity Index (SSIM)\cite{ssim} (calculated with $7\times7$\,pixel uniform windows, only considering windows fully inside the brain), PINQI achieves the best, i.e.\ highest, result of $0.939$ compared to the other methods.}
Examples of parameter maps for $T_1$ as well as the magnitude of $M_0$ obtained for a random slice of the test dataset are shown in \Fref{fig:result_silico}. Here, only PINQI was able to resolve the fine details.
\rev{At lower undersampling factors, PINQI also achieves superior results compared to all comparison methods. For example, at 4-fold undersampling it yields $0.073/0.042/0.961$ in terms of nRMSE/MAE/SSIM compared to the second best results of $0.091/0.054/0.938$.}

\subsection{Ablation Study}
\begin{figure}
    \centering
    \includegraphics[width=1.03\onecol]{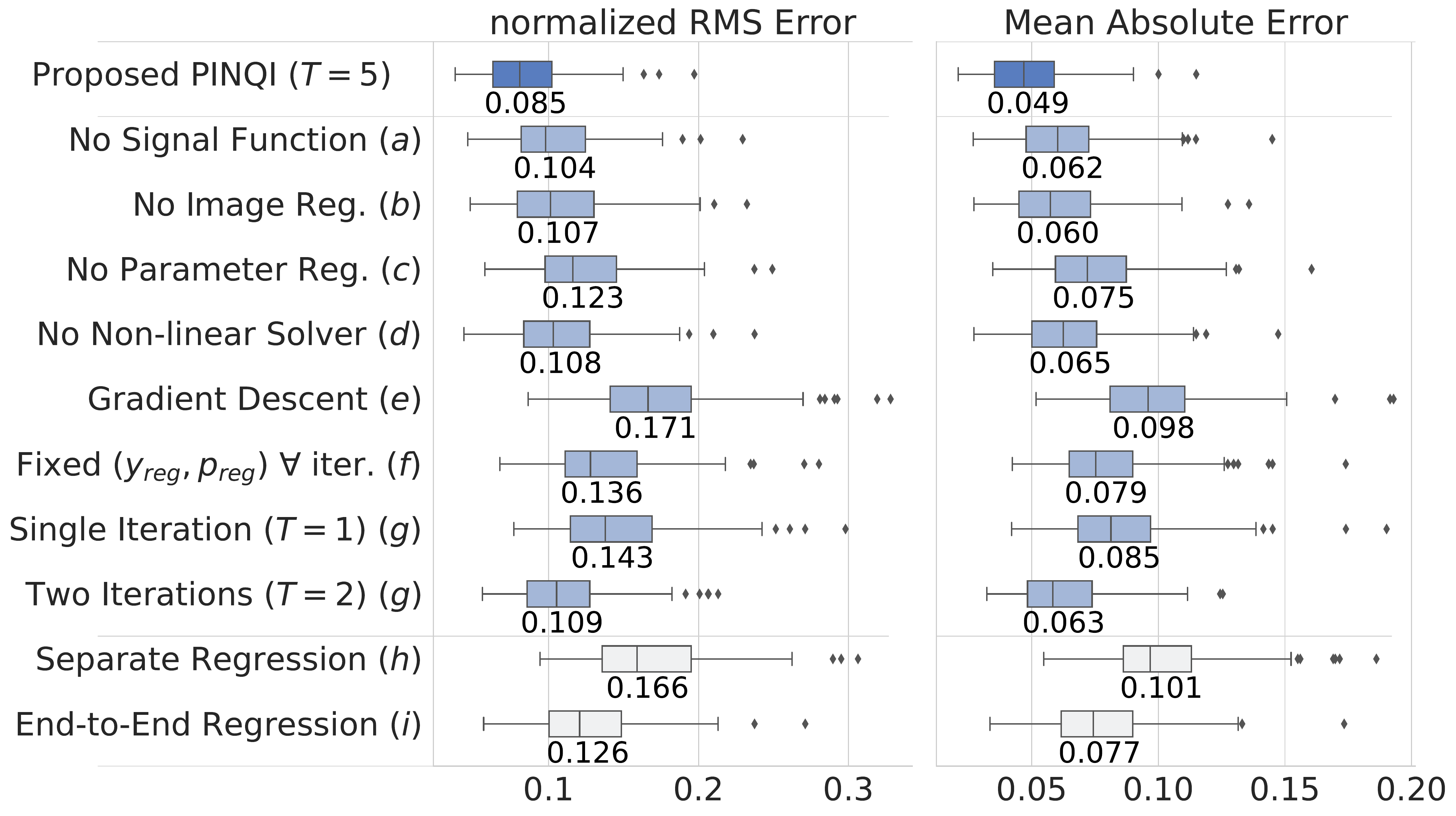}
    \caption{Results of the ablation study for $T_1$, highlighting the importance of iteratively applying both regularizing UNets as well the inclusion of the signal function via a non-linear regression layer. Note, here $N_r = 8$ and the saturation times $\tau$ differ from those used in \Fref{fig:result_silico_box}. \rev{For further details, see the full descriptions of the experiments in  \Fref{sec:ablation} with corresponding captions.}}
    \label{fig:ablation}
\end{figure}

The performance results obtained from the ablations are summarized in terms of nRMSE and MAE of $T_1$ in \Fref{fig:ablation}. The greatest increase in both MAE and nRMSE compared to PINQI as proposed is observed if, \rev{instead of the proposed differentiable optimization layer, gradient descent steps with a learned step size is used within the network (\Fref{fig:ablation}, \textit{e)}). This showcases the importance of the proposed optimization layer for PINQI. The next biggest increases are if either the UNets are only applied once to obtain $\vec{p}_{\mathrm{reg}}$ and $\vec{y}_{\mathrm{reg}}$ independent of the iteration of the unrolled scheme (\Fref{fig:ablation}, \textit{f)}), the iteration number is drastically decreased to $T=1$ (\Fref{fig:ablation}, \textit{g)}), or the parameter regularization network is completely removed (\Fref{fig:ablation}, \textit{b)}).} These observations highlight the importance of learned regularization. The removal of the L-BFGS solver in the non-linear subproblem \ref{eq:sub_non-linear_F}, corresponding to setting $\lambda_p^i = \infty$, resulting in $\vec{p}^i=\vec{p}^i_{_\theta}$, also severely degrades the performance. In this ablation, the composition of parameter estimation UNet and $q$ can be interpreted as a learned proximal mapping, \rev{as used in other unrolled reconstruction methods\cite{deepunrolling}}. We observed further degradation by full removal of the explicit knowledge about the signal function $q$, i.e.\ only solving the linear image reconstruction problem with Fourier-space data-consistency and using a UNet for parameter prediction, \rev{ similar as in other recent methods \cite{deept1,correct}.}

The comparison between a learned reconstruction of qualitative images with the \rev{\textit{separate}} parameter regression as a  post-processing step (\Fref{fig:ablation}, \textit{h)}), and the inclusion of the regression in the network with \textit{end-to-end} training \rev{utilizing the proposed differentiable optimization layer (\Fref{fig:ablation}, \textit{i)}),} shows a reduction of the mean nRMSE of the $T_1$-maps of ${>}20\%$.

\subsection{Phantom and In-Vivo Application}
We applied the final trained network to the data acquired from the physical phantom and  used the reference measurement to obtain the mean $T_1$ for each region-of-interest (ROI) for comparison. The RMS deviation between our proposed method and these reference values was \rev{35}\, ms, the full results for all nine ROIs are shown in \Fref{tab:phantom}.

\begin{table}
\caption{$T_1$-values (in s) and standard deviations over ROIs obtained for the physical phantom by the fully-sampled reference method with 18 saturations delays compared to PINQI with 8-fold undersampling and 5 delays.}

    \centering
    \setlength{\tabcolsep}{4pt}
    \begin{tabular}{r|r r r r r r r r r}
    \hline
         Reference&0.28 & 0.38 & 0.39 & 0.45 & 0.59 & 0.60 & 1.15 & 1.42 & 1.76\\
         Std. Dev. &0.01 & 0.01 & 0.03 & 0.01 & 0.02 & 0.02 & 0.04 & 0.06 & 0.13\\
         \hline
     PINQI  & \rev{0.29} & \rev{0.37} & \rev{0.42} & \rev{0.46} & \rev{0.60} & \rev{0.59} & \rev{1.10} & \rev{1.47} & \rev{1.76}\\
         
        Std. Dev. &\rev{0.03} & \rev{0.02} & \rev{0.06} &\rev{ 0.05} &\rev{ 0.02} & \rev{0.02} & \rev{0.05} &\rev{ 0.16} &\rev{ 0.14}\\
         \hline
         Difference & 4 \% & -1 \% & 7 \% & 3 \% & 2 \% & -2 \% & -4 \% & 4 \% &1 \%\\
         \hline
    \end{tabular}
    \label{tab:phantom}
\end{table}

\begin{figure}
    \centering
    \includegraphics[width=\onecol]{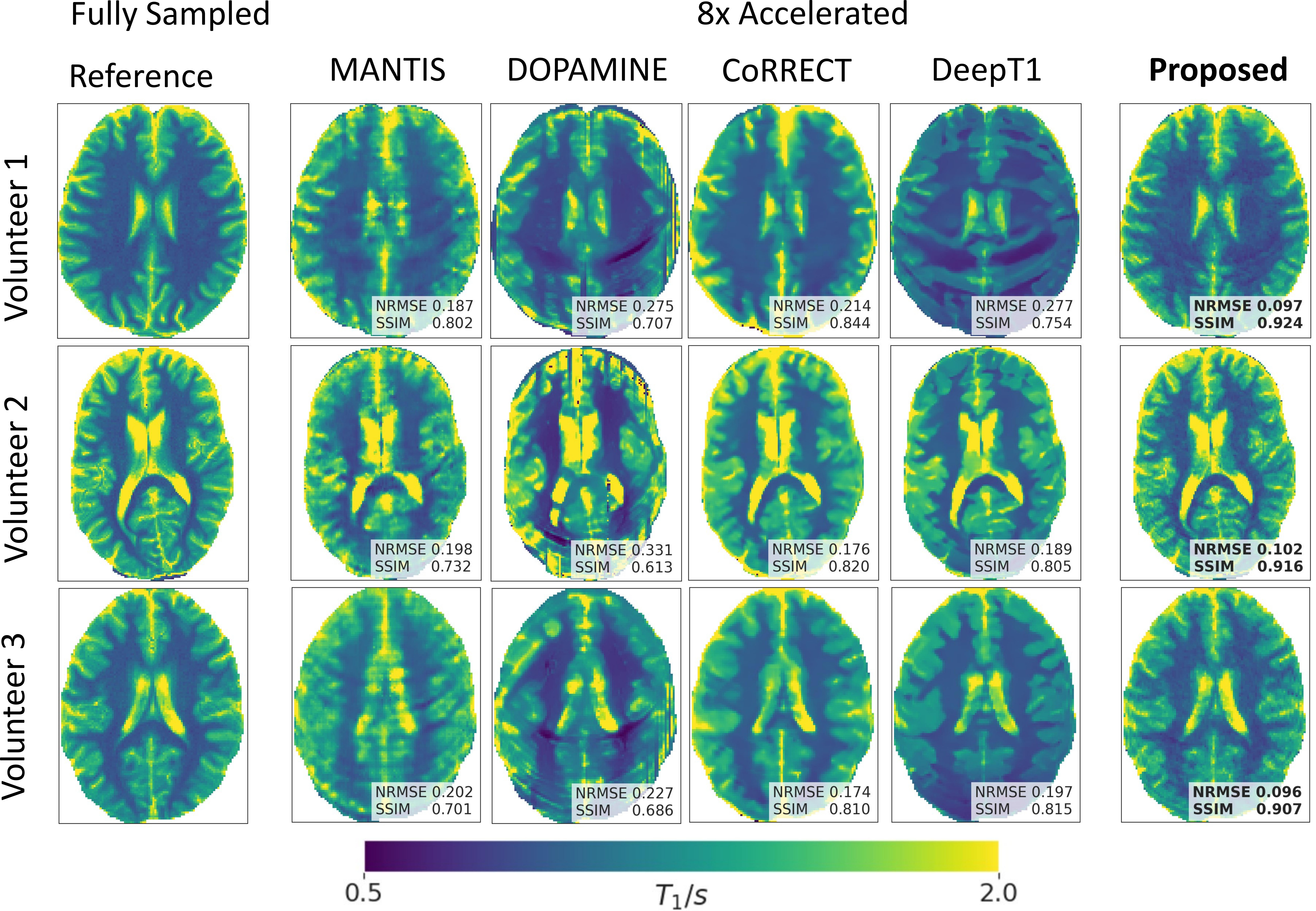}
    \caption{Examples of application of our proposed PINQI and comparison methods to 8-fold accelerated saturation recovery scans of the brains of three volunteers. All networks were solely trained on synthetic data. Only PINQI can successfully remove most of the artifacts caused by the severe undersampling and results in $T_1$ closely matching the fully sampled saturation recovery measurements used as reference.}
    \label{fig:invivo}
\end{figure}

To qualitatively demonstrate the applicability of the proposed method to real, unseen in-vivo measurements, we present the results of the volunteer study in \Fref{fig:invivo}. Even though the acquisition was severely undersampled, the network predicts $T_1$-maps in agreement with the fully-sampled references and only minor artifacts remain.

\section{Discussion}
We have demonstrated the superiority of PINQI for saturation recovery $T_1$-mapping compared to \rev{four} current state-of-the-art qMRI reconstruction techniques, as evidenced by the lower $T_1$-errors on the synthetic test dataset (\Fref{fig:result_silico_box}). Given that only $T_1$ holds clinical significance among the parameters obtainable in a saturation recovery sequence, and considering that most of the compared methods do not provide access to the phase of $M_0$,  our quantitative comparison here was primarily focused on $T_1$. 
The methods used for comparison, MANTIS, DeepT1, \rev{CoRRECT}, and DOPAMINE all have their own characteristics and limitations. MANTIS utilizes a fully learned mapping from magnitude qualitative images with artifacts to quantitative parameter maps, but only incorporates the physical model during training. DeepT1 \rev{and CoRRECT} enforce data-consistency between qualitative images and recorded k-space data, but do not enforce consistency between the final predicted parameter maps and recorded data at inference time. DOPAMINE explicitly uses knowledge about the non-linear signal model during inference, but does not employ image-space regularization and can use only shallow CNNs for parameter-space regularization due to the high number of iterations necessary.
In contrast, PINQI incorporates data-consistency for the non-linear signal model through non-linear optimization layers, allowing for the inclusion of prior knowledge about the underlying physics at each iteration of the unrolled reconstruction. Additionally, PINQI makes use of both image- and parameter-space regularization and explicitly formulates the influence of all UNets as a learned regularizer.
In total, these distinctions position PINQI as a unique method in qMRI reconstruction.
Note that for a fair comparison, we reimplemented and adjusted the comparison methods to our dataset and signal model, as well as optimized hyperparameters rather than relying on the choices of the respective works. We were unable to use DOPAMINE as published, as training was highly unstable. We speculate, that the combination of an iteration-dependent but data-independent stepsize factor and a first-order gradient step without any thresholds was causing high susceptibility to even minor outliers. We had to modify the training to overcome this issue.
\rev{
During training with a batch size of 8, our PyTorch implementation of PINQI utilizes approximately 36\,GB of memory in our specific configuration. This is comparable to our implementations of CoRRECT (34\,GB), DeepT1 (23\,GB), and DOPAMINE (17\,GB), but significantly higher than MANTIS (2\,GB). It is crucial to note that these figures are highly dependent on the implementation details and the nature of the problem. Therefore, they should be interpreted as rough guidance for estimating memory consumption.}

In our ablation study, we highlighted the importance of both image- and parameter-space regularization networks, as well as the utilization of the non-linear optimization layer and the signal model. Removal of any of these major components resulted in severe degradation, as summarized in \Fref{fig:ablation}. By employing an unrolling approach and alternating between the two subproblems, we were able to update the predictions used as learned regularizations in each iteration, which proved to be highly beneficial compared to a single application of the UNets. This suggests, that the UNets were able to make use of the gradually restored information within the unrolled algorithm. 
Moreover, we have demonstrated that the addition of a single non-linear optimization, as a final layer, to an otherwise unchanged learned reconstruction, can significantly reduce the error in the resulting parameter maps compared to performing a separate regression as a post-processing step outside the network (see  \Fref{fig:ablation}, \textit{End-to-End Regression (i)} vs.\ \textit{Separate Regression (h)}). This approach can be seamlessly integrated into existing reconstruction methods, enabling training with an objective that aligns more closely with the true objective in quantitative MRI. Essentially, it can be viewed as a task-specific loss function, which effectively penalizes reconstruction errors of the intermediate qualitative images that contribute to poor final regression results. Unlike a simple $L_p$ loss on the qualitative intermediate images, this approach allows for more efficient learning by focusing on important features \cite{cramerloss,kendallmulti}. We emphasize the potential of adding a single differentiable regression layer to both existing and future quantitative imaging methods as a straightforward means of incorporating knowledge about the signal model. 
\rev{As the implicit differentiation-based numerical gradient calculation only provides an approximation in a neighborhood of the possibly local, minimum obtained by the inner optimization, the usage may necessitate additional regularization, as employed in PINQI. Additionally, gradient noise might be an issue for certain problems.}

Finally, we demonstrated that PINQI, even when trained solely on synthetic data, can be successfully applied to real data. This was confirmed both quantitatively, with low deviations from the reference measurements of the physical phantom in \Fref{tab:phantom}, and qualitatively, through the reconstruction of nearly artifact-free maps from undersampled data in \Fref{fig:invivo}. In contrast, the comparison methods cannot faithfully reconstruct the parameter maps in this setting with significantly higher undersampling than in the original publications and without specifically acquired in-vivo training data.
\rev{In our synthetic evaluation, the data-consistent reconstruction methods were able to utilize the ground truth forward model, specifically $q$ and the true sensitivity maps used the acquisition operator $\op{A}$, whereas in the tests on real scanner data, these are not available.
Here, we also acknowledge} the possibility of self-supervised fine-tuning of our method for adaptation from the assumed forward model during training to the partially unknown and potentially different true forward model during testing, \rev{or a shifted distribution of the parameter maps \cite{adaptation,systemic,ismrm,ssl}}. In particular, the sensitivity maps estimated from undersampled data could be enhanced by an additional subnetwork \cite{sriram}. However, for the sake of simplicity and comparability, we refrained from implementing such modules in our network or in any of the comparison methods. Nevertheless, \Fref{eq:csmgrad}, obtained through implicit differentiation, would facilitate an efficient extension.

\rev{Although the formulation of an optimization objective with learned regularization provides some interpretability compared to a direct learned mapping of the quantitative parameters, as in, for example, CoRRECT, MANTIS, or DeepT1, the inner workings of PINQI remain largely a black-box. The unrolling, the repeated application of the subnetworks, and the inner optimizers increase the memory requirements during training as well as the computational requirements at inference time of PINQI compared to the comparison methods, potentially limiting its application in settings without suitable accelerator hardware or in 3D acquisitions. Notably, in our synthetic setting, we demonstrated that the results of the proposed PINQI at an acceleration factor of 8 are superior to those obtained by all other comparison methods at 4-fold acceleration. Despite the increased reconstruction time (2.3\,s for 8 slices, using an Nvidia A6000) compared to, for example, DeepT1 (0.6\,s) and CoRRECT (0.3\,s), opting for faster acquisition at lower errors can be an acceptable trade-off. Finally, PINQI requires a known, twice-differentiable closed-form signal model, which may limit its direct application to techniques such as magnetic resonance fingerprinting.
}

While we have only demonstrated the application of PINQI to Cartesian undersampled $T_1$-mapping, the splitting of PINQI can be trivially adapted to many inverse problems of the form described by \Fref{eq:problem} by adjusting the non-linear signal model and the linear acquisition operator. Thus, \rev{while its effectiveness for other signal models remains to be further explored,} PINQI represents a general physics-informed network for many quantitative imaging inverse problems.

\section{Conclusion}
Our proposed method, PINQI, presents a novel approach to quantitative image reconstruction by combining unrolled optimization, differentiable optimization layers, and learned regularization to fully utilize prior knowledge regarding the underlying physics.
On a representative qMRI task, $T_1$ mapping of the brain, we demonstrated superiority compared to three established data-driven reconstruction methods on a synthetic test dataset. Finally, we showed that our model, while trained on synthetic data alone, is transferable to in-vivo data.

\bibliography{bib2}

\begin{thebibliography}{10}
\providecommand{\url}[1]{#1}
\csname url@samestyle\endcsname
\providecommand{\newblock}{\relax}
\providecommand{\bibinfo}[2]{#2}
\providecommand{\BIBentrySTDinterwordspacing}{\spaceskip=0pt\relax}
\providecommand{\BIBentryALTinterwordstretchfactor}{4}
\providecommand{\BIBentryALTinterwordspacing}{\spaceskip=\fontdimen2\font plus
\BIBentryALTinterwordstretchfactor\fontdimen3\font minus
  \fontdimen4\font\relax}
\providecommand{\BIBforeignlanguage}[2]{{%
\expandafter\ifx\csname l@#1\endcsname\relax
\typeout{** WARNING: IEEEtran.bst: No hyphenation pattern has been}%
\typeout{** loaded for the language `#1'. Using the pattern for}%
\typeout{** the default language instead.}%
\else
\language=\csname l@#1\endcsname
\fi
#2}}
\providecommand{\BIBdecl}{\relax}
\BIBdecl

\bibitem{sense}
K.~P. Pruessmann, M.~Weiger, M.~B. Scheidegger, and P.~Boesiger, ``{SENSE:
  Sensitivity encoding for fast MRI},'' \emph{Magnetic Resonance in Medicine},
  vol.~42, no.~5, pp. 952--962, 1999.
  10.1002/(SICI)1522-2594(199911)42:5<952::AID-MRM16>3.0.CO;2-S

\bibitem{lustigcs}
M.~Lustig, D.~L. Donoho, J.~M. Santos, and J.~M. Pauly, ``{Compressed Sensing
  MRI},'' \emph{IEEE Signal Processing Magazine}, vol.~25, no.~2, pp. 72--82,
  2008. 10.1109/MSP.2007.914728

\bibitem{doneva2010}
M.~Doneva, P.~B{\"{o}}rnert, H.~Eggers, C.~Stehning, J.~S{\'{e}}n{\'{e}}gas,
  and A.~Mertins, ``{Compressed sensing reconstruction for magnetic resonance
  parameter mapping},'' \emph{Magnetic Resonance in Medicine}, vol.~64, no.~4,
  pp. 1114--1120, 2010. 10.1002/mrm.22483

\bibitem{wang2018}
X.~Wang, V.~Roeloffs, J.~Klosowski, Z.~Tan, D.~Voit, M.~Uecker, and J.~Frahm,
  ``{Model-based T1 mapping with sparsity constraints using single-shot
  inversion-recovery radial FLASH},'' \emph{Magnetic Resonance in Medicine},
  vol.~79, no.~2, pp. 730--740, 2018. 10.1002/mrm.26726

\bibitem{zhao2014}
B.~Zhao, F.~Lam, and Z.~P. Liang, ``{Model-based MR parameter mapping with
  sparsity constraints: Parameter estimation and performance bounds},''
  \emph{IEEE Transactions on Medical Imaging}, vol.~33, no.~9, pp. 1832--1844,
  2014. 10.1109/TMI.2014.2322815

\bibitem{Block2007}
K.~T. Block, M.~Uecker, and J.~Frahm, ``{Undersampled radial MRI with multiple
  coils. Iterative image reconstruction using a total variation constraint},''
  \emph{Magnetic Resonance in Medicine}, vol.~57, no.~6, pp. 1086--1098, 2007.
  10.1002/mrm.21236

\bibitem{tran2013}
J.~Tran-Gia, D.~St{\"{a}}b, T.~Wech, D.~Hahn, and H.~K{\"{o}}stler,
  ``{Model-based Acceleration of Parameter mapping (MAP) for saturation
  prepared radially acquired data},'' \emph{Magnetic Resonance in Medicine},
  vol.~70, no.~6, pp. 1524--1534, 2013. 10.1002/mrm.24600

\bibitem{becker2019}
K.~M. Becker, J.~Schulz-Menger, T.~Schaeffter, and C.~Kolbitsch,
  ``{Simultaneous high-resolution cardiac T1 mapping and cine imaging using
  model-based iterative image reconstruction},'' \emph{Magnetic Resonance in
  Medicine}, vol.~81, no.~2, pp. 1080--1091, 2019. 10.1002/mrm.27474

\bibitem{sriram}
A.~Sriram, J.~Zbontar, T.~Murrell, A.~Defazio, C.~L. Zitnick, N.~Yakubova,
  F.~Knoll, and P.~Johnson, ``{End-to-End Variational Networks for Accelerated
  MRI Reconstruction},'' \emph{Medical Image Computing and Computer Assisted
  Intervention – MICCAI 2020. Lecture Notes in Computer Science}, vol. 12262
  LNCS, pp. 64--73, 2020. 10.1007/978-3-030-59713-9{\_}7

\bibitem{modl}
H.~K. Aggarwal, M.~P. Mani, and M.~Jacob, ``{MoDL: Model-Based Deep Learning
  Architecture for Inverse Problems},'' \emph{IEEE Transactions on Medical
  Imaging}, vol.~38, no.~2, pp. 394--405, 2019. 10.1109/TMI.2018.2865356

\bibitem{dopamine}
Y.~Jun, H.~Shin, T.~Eo, T.~Kim, and D.~Hwang, ``{Deep model-based magnetic
  resonance parameter mapping network (DOPAMINE) for fast T1 mapping using
  variable flip angle method},'' \emph{Medical Image Analysis}, vol.~70, p.
  102017, 2021. 10.1016/j.media.2021.102017

\bibitem{deept1}
H.~Jeelani, Y.~Yang, R.~Zhou, C.~M. Kramer, M.~Salerno, and D.~S. Weller, ``{A
  Myocardial T1-Mapping Framework with Recurrent and U-Net Convolutional Neural
  Networks},'' \emph{Proceedings - International Symposium on Biomedical
  Imaging}, vol. 2020-April, pp. 1941--1944, 2020.
  10.1109/ISBI45749.2020.9098459

\bibitem{correct}
X.~Xu, W.~Gan, S.~V. Kothapalli, D.~A. Yablonskiy, and U.~S. Kamilov,
  ``{CoRRECT: A Deep Unfolding Framework for Motion-Corrected Quantitative R2*
  Mapping},'' 2022. 10.48550/arXiv.2210.06330

\bibitem{mantis}
F.~Liu, L.~Feng, and R.~Kijowski, ``{MANTIS: Model-Augmented Neural neTwork
  with Incoherent k-space Sampling for efficient MR parameter mapping},''
  \emph{Magnetic Resonance in Medicine}, vol.~82, no.~1, pp. 174--188, 2019.
  10.1002/mrm.27707

\bibitem{myomapnet}
R.~Guo, H.~El-Rewaidy, S.~Assana, X.~Cai, A.~Amyar, K.~Chow, X.~Bi, T.~Yankama,
  J.~Cirillo, P.~Pierce, B.~Goddu, L.~Ngo, and R.~Nezafat, ``{Accelerated
  cardiac T1 mapping in four heartbeats with inline MyoMapNet: a deep
  learning-based T1 estimation approach},'' \emph{Journal of Cardiovascular
  Magnetic Resonance}, vol.~24, no.~1, pp. 1--15, 2022.
  10.1186/s12968-021-00834-0

\bibitem{drone}
O.~Cohen, B.~Zhu, and M.~S. Rosen, ``{MR fingerprinting Deep RecOnstruction
  NEtwork (DRONE)},'' \emph{Magnetic Resonance in Medicine}, vol.~80, no.~3,
  pp. 885--894, 2018. 10.1002/mrm.27198

\bibitem{schlemper}
J.~Schlemper, J.~Caballero, J.~V. Hajnal, A.~N. Price, and D.~Rueckert, ``{A
  Deep Cascade of Convolutional Neural Networks for Dynamic MR Image
  Reconstruction},'' \emph{IEEE Transactions on Medical Imaging}, vol.~37,
  no.~2, pp. 491--503, 2018. {10.1007/978-3-319-59050-9_51}

\bibitem{crnn}
C.~Qin, J.~Schlemper, J.~Caballero, A.~N. Price, J.~V. Hajnal, and D.~Rueckert,
  ``{Convolutional recurrent neural networks for dynamic MR image
  reconstruction},'' \emph{IEEE Transactions on Medical Imaging}, vol.~38,
  no.~1, pp. 280--290, 2019. 10.1109/TMI.2018.2863670

\bibitem{admmnet}
Y.~Yang, J.~Sun, H.~Li, and Z.~Xu, ``{Deep ADMM-Net for compressive sensing
  MRI},'' \emph{Advances in Neural Information Processing Systems}, vol.~29,
  2016. 10.5555/3157096.3157098

\bibitem{sasha}
K.~Chow, J.~A. Flewitt, J.~D. Green, J.~J. Pagano, M.~G. Friedrich, and R.~B.
  Thompson, ``{Saturation recovery single-shot acquisition (SASHA) for
  myocardial T 1 mapping},'' \emph{Magnetic Resonance in Medicine}, vol.~71,
  no.~6, pp. 2082--2095, 2014. 10.1002/mrm.24878

\bibitem{bojorquez2017}
J.~Z. Bojorquez, S.~Bricq, C.~Acquitter, F.~Brunotte, P.~M. Walker, and
  A.~Lalande, ``{What are normal relaxation times of tissues at 3 T?}''
  \emph{Magnetic Resonance Imaging}, vol.~35, pp. 69--80, 2017.
  10.1016/j.mri.2016.08.021

\bibitem{bfgs}
D.~C. Liu and J.~Nocedal, ``{On the limited memory BFGS method for large scale
  optimization},'' \emph{Mathematical Programming}, vol.~45, no. 1-3, pp.
  503--528, 8 1989. 10.1007/BF01589116

\bibitem{numopt}
J.~Nocedal and S.~Wright, \emph{{Numerical Optimization}}, ser. Springer Series
  in Operations Research and Financial Engineering.\hskip 1em plus 0.5em minus
  0.4em\relax Springer New York, 2006. ISBN 9780387303031

\bibitem{mrf}
D.~Ma, V.~Gulani, N.~Seiberlich, K.~Liu, J.~L. Sunshine, J.~L. Duerk, and M.~A.
  Griswold, ``{Magnetic resonance fingerprinting},'' \emph{Nature}, vol. 495,
  no. 7440, pp. 187--192, 3 2013. 10.1038/nature11971

\bibitem{supermap}
H.~Li, M.~Yang, J.~H. Kim, C.~Zhang, R.~Liu, P.~Huang, D.~Liang, X.~Zhang,
  X.~Li, and L.~Ying, ``{SuperMAP: Deep ultrafast MR relaxometry with joint
  spatiotemporal undersampling},'' \emph{Magnetic Resonance in Medicine},
  vol.~89, no.~1, pp. 64--76, 2023. 10.1002/mrm.29411

\bibitem{ronneberger2015}
O.~Ronneberger, P.~Fischer, and T.~Brox, ``{U-net: Convolutional networks for
  biomedical image segmentation},'' \emph{Lecture Notes in Computer Science
  (including subseries Lecture Notes in Artificial Intelligence and Lecture
  Notes in Bioinformatics)}, vol. 9351, pp. 234--241, 2015.
  10.1007/978-3-319-24574-4{\_}28

\bibitem{hammernikreview}
K.~Hammernik, T.~Kustner, B.~Yaman, Z.~Huang, D.~Rueckert, F.~Knoll, and
  M.~Akcakaya, ``{Physics-Driven Deep Learning for Computational Magnetic
  Resonance Imaging: Combining physics and machine learning for improved
  medical imaging},'' \emph{IEEE Signal Processing Magazine}, vol.~40, no.~1,
  pp. 98--114, 2023. 10.1109/msp.2022.3215288

\bibitem{deepunrolling}
D.~Chen, M.~E. Davies, and M.~Golbabaee, ``{Deep Unrolling for Magnetic
  Resonance Fingerprinting},'' \emph{Proceedings - International Symposium on
  Biomedical Imaging}, no.~2, 2022. 10.1109/ISBI52829.2022.9761475

\bibitem{kofler}
A.~Kofler, M.~Haltmeier, T.~Schaeffter, and C.~Kolbitsch, ``An
  end-to-end-trainable iterative network architecture for accelerated radial
  multi-coil 2d cine mr image reconstruction,'' \emph{Medical Physics},
  vol.~48, no.~5, pp. 2412--2425, 2021.

\bibitem{Amos2017}
B.~Amos and J.~Z. Kolter, ``{OptNet: Differentiable optimization as a layer in
  neural networks},'' \emph{34th International Conference on Machine Learning,
  ICML 2017}, vol.~1, pp. 179--191, 2017. 10.48550/arXiv.1703.0044

\bibitem{Agrawal2019}
A.~Agrawal, B.~Amos, S.~Barratt, S.~Boyd, S.~Diamond, and J.~Zico~Kolter,
  ``{Differentiable convex optimization layers},'' \emph{Advances in Neural
  Information Processing Systems}, vol.~32, no. NeurIPS, 2019.
  10.48550/arXiv.1910.12430

\bibitem{diffcomp}
Z.~Lv, F.~Dellaert, J.~M. Rehg, and A.~Geiger, ``{Taking a deeper look at the
  inverse compositional algorithm},'' \emph{Proceedings of the IEEE Computer
  Society Conference on Computer Vision and Pattern Recognition}, vol.
  2019-June, pp. 4576--4585, 2019. 10.1109/CVPR.2019.00471

\bibitem{hammernikbook}
K.~Hammernik, T.~K{\"{u}}stner, and D.~Rueckert, ``{Machine Learning for MRI
  Reconstruction},'' in \emph{Magnetic Resonance Image Reconstruction},
  C.~Prieto, M.~I. Doneva, and M.~Akcakaya, Eds.\hskip 1em plus 0.5em minus
  0.4em\relax Elsevier, 2022, ch.~11, pp. 281--317.

\bibitem{Gould2016}
S.~Gould, B.~Fernando, A.~Cherian, P.~Anderson, R.~S. Cruz, and E.~Guo, ``On
  differentiating parameterized argmin and argmax problems with application to
  bi-level optimization,'' 2016.

\bibitem{crowder}
M.~Crowder, J.~R. Magnus, and H.~Neudecker, \emph{{Matrix Differential Calculus
  with Applications in Statistics and Econometrics.}}\hskip 1em plus 0.5em
  minus 0.4em\relax John Wiley \& Sons, 1989. ISBN 9781119541202

\bibitem{Oliveira2013}
O.~de~Oliveira, ``{The implicit and the inverse function theorems: Easy
  proofs},'' \emph{Real Analysis Exchange}, vol.~39, no.~1, pp. 207--218, 2013.
  10.14321/realanalexch.39.1.0207

\bibitem{pedregosa}
F.~Pedregosa, ``{Hyperparameter optimization with approximate gradient},''
  \emph{33rd International Conference on Machine Learning, ICML 2016}, vol.~2,
  pp. 1150--1159, 2016. 10.48550/arXiv.1602.02355

\bibitem{Afonso2010}
M.~V. Afonso, J.~M. Bioucas-Dias, and M.~A. Figueiredo, ``{Fast image recovery
  using variable splitting and constrained optimization},'' \emph{IEEE
  Transactions on Image Processing}, vol.~19, no.~9, pp. 2345--2356, 2010.
  10.1109/TIP.2010.2047910

\bibitem{practical}
P.~E. Gill, W.~Murray, and M.~H. Wright, \emph{{Practical Optimization}}.\hskip
  1em plus 0.5em minus 0.4em\relax Academic Press, 1981. ISBN 9780122839504

\bibitem{wirtinger}
K.~Kreutz-Delgado, ``{The complex gradient operator and the CR-calculus},''
  \emph{arXiv preprint}, 2009. 10.48550/arXiv.0906.4835

\bibitem{brainweb}
B.~Aubert-Broche, M.~Griffin, G.~B. Pike, A.~C. Evans, and D.~L. Collins,
  ``{Twenty new digital brain phantoms for creation of validation image data
  bases},'' \emph{IEEE Transactions on Medical Imaging}, vol.~25, no.~11, pp.
  1410--1416, 2006. 10.1109/TMI.2006.883453

\bibitem{torchkbnufft}
M.~J. Muckley, R.~Stern, T.~Murrell, and F.~Knoll, ``mrisensesim.py,''
  {TorchKbNufft}: A High-Level, Hardware-Agnostic Non-Uniform Fast {Fourier}
  Transform, ISMRM Workshop on Data Sampling \& Image Reconstruction. Online
  \href{https://github.com/mmuckley/torchkbnufft/tree/v0.3.4/torchkbnufft/mri}{github.com/mmuckley/torchkbnufft/tree/v0.3.4/torchkbnufft/mri}.

\bibitem{walsh}
D.~O. Walsh, A.~F. Gmitro, and M.~W. Marcellin, ``{Adaptive reconstruction of
  phased array MR imagery},'' \emph{Magnetic Resonance in Medicine}, vol.~43,
  no.~5, pp. 682--690, 2000.
  10.1002/(SICI)1522-2594(200005)43:5<682::AID-MRM10>3.0.CO;2-G

\bibitem{inati}
S.~J. Inati, M.~S. Hansen, and P.~Kellman, ``{A Fast Optimal Method for Coil
  Sensitivity Estimation and Adaptive Coil Combination for Complex Images},''
  \emph{Proceedings of the 22nd Annual Meeting of ISMRM}, 2014.

\bibitem{espirit}
M.~Uecker, P.~Lai, M.~J. Murphy, P.~Virtue, M.~Elad, J.~M. Pauly, S.~S.
  Vasanawala, and M.~Lustig, ``{ESPIRiT}--an eigenvalue approach to
  autocalibrating parallel {MRI}: where {SENSE} meets {GRAPPA},''
  \emph{Magnetic Resonance in Medicine}, vol.~71, no.~3, pp. 990--1001, 2014.
  10.1002/mrm.24751

\bibitem{pulseq}
K.~J. Layton, S.~Kroboth, F.~Jia, S.~Littin, H.~Yu, J.~Leupold, J.-F. Nielsen,
  T.~Stöcker, and M.~Zaitsev, ``Pulseq: A rapid and hardware-independent pulse
  sequence prototyping framework,'' \emph{Magnetic Resonance in Medicine},
  vol.~77, no.~4, pp. 1544--1552, 2017. 10.1002/mrm.26235

\bibitem{silu}
B.~Zoph and Q.~V. Le, ``{Searching for activation functions},'' \emph{6th
  International Conference on Learning Representations, ICLR 2018 - Workshop
  Track Proceedings}, no.~1, pp. 1--12, 2018. 10.48550/arXiv.1710.05941

\bibitem{groupnorm}
Y.~Wu and K.~He, ``{Group Normalization},'' \emph{International Journal of
  Computer Vision}, vol. 128, no.~3, pp. 742--755, 2020.
  10.1007/s11263-019-01198-w

\bibitem{Qiu2017}
Z.~Qiu, T.~Yao, and T.~Mei, ``{Learning Spatio-Temporal Representation with
  Pseudo-3D Residual Networks},'' \emph{Proceedings of the IEEE International
  Conference on Computer Vision}, vol. 2017-Octob, pp. 5534--5542, 10 2017.
  10.1109/ICCV.2017.590

\bibitem{film}
E.~Perez, F.~Strub, H.~De~Vries, V.~Dumoulin, and A.~Courville, ``{FiLM: Visual
  reasoning with a general conditioning layer},'' \emph{32nd AAAI Conference on
  Artificial Intelligence, AAAI 2018}, pp. 3942--3951, 2018.
  10.1609/aaai.v32i1.11671

\bibitem{nosense}
F.~F. Zimmermann and A.~Kofler, ``{NoSENSE: Learned unrolled cardiac MRI
  reconstruction without explicit sensitivity maps},'' \emph{International
  Workshop on Statistical Atlases and Computational Models of the Heart
  (STACOM)}, 2023. 10.48550/arXiv.2309.15608

\bibitem{rezero}
T.~Bachlechner, B.~P. Majumder, H.~Mao, G.~Cottrell, and J.~McAuley, ``{ReZero
  is All You Need: Fast Convergence at Large Depth},'' \emph{37th Conference on
  Uncertainty in Artificial Intelligence, UAI 2021}, no.~1, pp. 1352--1361,
  2021. 10.48550/arXiv./2003.04887

\bibitem{dudornet}
B.~Zhou and S.~Kevin~Zhou, ``{Dudornet: Learning a dual-domain recurrent
  network for fast MRI reconstruction with deep T1 prior},'' \emph{Proceedings
  of the IEEE Computer Society Conference on Computer Vision and Pattern
  Recognition}, pp. 4272--4281, 2020. 10.1109/CVPR42600.2020.00433

\bibitem{deepsupervision}
L.~Wang, C.-Y. Lee, Z.~Tu, and S.~Lazebnik, ``Training deeper convolutional
  networks with deep supervision,'' 2015.

\bibitem{adam}
D.~P. Kingma and J.~L. Ba, ``{Adam: A method for stochastic optimization},''
  \emph{3rd International Conference on Learning Representations, ICLR 2015 -
  Conference Track Proceedings}, pp. 1--15, 2015. 10.48550/arXiv.412.6980

\bibitem{he2019}
T.~He, Z.~Zhang, H.~Zhang, Z.~Zhang, J.~Xie, and M.~Li, ``{Bag of tricks for
  image classification with convolutional neural networks},'' \emph{Proceedings
  of the IEEE Computer Society Conference on Computer Vision and Pattern
  Recognition}, vol. 2019-June, pp. 558--567, 2019. 10.48550/arXiv.1812.01187

\bibitem{adamw}
I.~Loshchilov and F.~Hutter, ``{Decoupled weight decay regularization},''
  \emph{7th International Conference on Learning Representations, ICLR 2019},
  2019. 10.48550/arXiv.1711.05101

\bibitem{ssim}
Z.~Wang, A.~C. Bovik, H.~R. Sheikh, and E.~P. Simoncelli, ``Image quality
  assessment: from error visibility to structural similarity,'' \emph{IEEE
  transactions on image processing}, vol.~13, no.~4, pp. 600--612, 2004.

\bibitem{cramerloss}
X.~Zhang, Q.~Duchemin, K.~Liu*, C.~Gultekin, S.~Flassbeck, C.~Fernandez-Granda,
  and J.~Assl{\"{a}}nder, ``{Cram{\'{e}}r–Rao bound-informed training of
  neural networks for quantitative MRI},'' \emph{Magnetic Resonance in
  Medicine}, vol.~88, no.~1, pp. 436--448, 2022. 10.1002/mrm.29206

\bibitem{kendallmulti}
A.~Kendall, Y.~Gal, and R.~Cipolla, ``Multi-task learning using uncertainty to
  weigh losses for scene geometry and semantics,'' in \emph{Proceedings of the
  IEEE conference on computer vision and pattern recognition}, 2018.
  10.48550/arXiv.1705.07115 pp. 7482--7491.

\bibitem{adaptation}
D.~Gilton, G.~Ongie, and R.~Willett, ``{Model Adaptation for Inverse Problems
  in Imaging},'' \emph{IEEE Transactions on Computational Imaging}, vol.~7,
  no.~2, pp. 661--674, 2021. 10.1109/TCI.2021.3094714

\bibitem{systemic}
K.~Hammernik, J.~Schlemper, C.~Qin, J.~Duan, R.~M. Summers, and D.~Rueckert,
  ``Systematic evaluation of iterative deep neural networks for fast parallel
  mri reconstruction with sensitivity-weighted coil combination,''
  \emph{Magnetic Resonance in Medicine}, vol.~86, no.~4, pp. 1859--1872, 2021.
  10.1002/mrm.28827

\bibitem{ismrm}
F.~F. Zimmermann, A.~Kofler, C.~Kolbitsch, and P.~Schuenke, ``{Semi-Supervised
  Learning for Spatially Regularized Quantitative MRI Reconstruction -
  Application to Simultaneous T1, B0, B1 Mapping},'' 2023, 1166, ISMRM Annual
  Meeting.

\bibitem{ssl}
Y.~Jun, J.~Cho, X.~Wang, M.~Gee, P.~E. Grant, B.~Bilgic, and B.~Gagoski,
  ``{SSL-QALAS: Self-Supervised Learning for rapid multiparameter estimation in
  quantitative MRI using 3D-QALAS},'' \emph{Magnetic Resonance in Medicine},
  vol.~90, no.~5, pp. 2019--2032, 2023. 10.1002/mrm.29786

\end{thebibliography}

\bibliographystyle{IEEEtran2}
\end{document}